\documentclass[reprint,
preprintnumbers,
nobibnotes,
superscriptaddress,
 amsmath,amssymb,
 aps,
prd,
]{revtex4-2}

\usepackage{graphicx, braket, orcidlink, algorithm2e, amssymb, amsmath, dsfont, adjustbox, multirow, lineno, nccmath, scalerel, xcolor, mathtools, booktabs}% Include figure files
\PassOptionsToPackage{breaklinks=true,colorlinks=true}{hyperref}
\PassOptionsToPackage{hyphens}{url}

\usepackage[normalem]{ulem}
\usepackage{dcolumn}% Align table columns on decimal point
\usepackage{bm}% bold math
\usepackage{tikz}
\usetikzlibrary{quantikz2}
\usepackage{physics} % Kyle added this, for his appendix. Hypothetically might collide undersirably with other packages (e.g. wherever you got your \ket from).

\hypersetup{
    unicode=true,
    pdftoolbar=true,
    pdfmenubar=true,
    pdffitwindow=false,
    pdfstartview={fitH},
    pdftitle={TEPID},
    pdfauthor={B.~Sambasivam, K.~Sherbert, K.~Shirali, N.J.~Mayhall, E.~Barnes, S.E.~Economou},
    pdfsubject={TEPID},
    pdfcreator={B.~Sambasivam, K.~Sherbert, K.~Shirali, N.J.~Mayhall, E.~Barnes, S.E.~Economou},
    pdfproducer={},
    pdfkeywords={},
    pdfnewwindow=true,
    colorlinks=true,
    linkcolor=blue,
    citecolor=magenta,
    filecolor=magenta,
    urlcolor=blue
}
\newcommand\Ueq{\stackrel{\mathclap{\scriptsize\mbox{$V$}}}{=}}
\newcommand\Uapprox{\stackrel{\mathclap{\scriptsize\mbox{$V$}}}{\approx}}
\DeclareMathOperator*{\argminA}{arg\,min} % Jan Hlavacek

\begin{document}

%\preprint{APS/TEPID-ADAPT}

\title{TEPID-ADAPT: Adaptive variational method for simultaneous preparation of low-temperature Gibbs and low-lying eigenstates
}% Force line breaks with \\

\author{Bharath Sambasivam\orcidlink{0000-0002-5765-9469}}
\email{sbharath@vt.edu}
\affiliation{Department of Physics, Virginia Tech, Blacksburg, VA 24061, USA}
\affiliation{Virginia Tech Center for Quantum Information Science and Engineering, Blacksburg, VA 24061, USA}

\author{Kyle Sherbert\orcidlink{0000-0002-5258-6539}}
%\email{kyle.sherbert@vt.edu}
\affiliation{Department of Physics, Virginia Tech, Blacksburg, VA 24061, USA}
\affiliation{Department of Chemistry, Virginia Tech, Blacksburg, VA 24061, USA}
\affiliation{Virginia Tech Center for Quantum Information Science and Engineering, Blacksburg, VA 24061, USA}

\author{Karunya Shirali\orcidlink{0000-0002-2006-2343}}
%\email{karunyashirali@vt.edu}
\affiliation{Department of Physics, Virginia Tech, Blacksburg, VA 24061, USA}
\affiliation{Virginia Tech Center for Quantum Information Science and Engineering, Blacksburg, VA 24061, USA}

\author{Nicholas J. Mayhall\orcidlink{0000-0002-1312-9781}}
%\email{nmayhall@vt.edu}
\affiliation{Department of Chemistry, Virginia Tech, Blacksburg, VA 24061, USA}
\affiliation{Virginia Tech Center for Quantum Information Science and Engineering, Blacksburg, VA 24061, USA}

\author{Aram W. Harrow\orcidlink{0000-0003-3220-7682}}
%\email{efbarnes@vt.edu}
\affiliation{Center for Theoretical Physics, Massachusetts Institute of Technology, Cambridge, MA 02139, USA}

\author{Edwin Barnes\orcidlink{0000-0002-0982-9339}}
%\email{efbarnes@vt.edu}
\affiliation{Department of Physics, Virginia Tech, Blacksburg, VA 24061, USA}
\affiliation{Virginia Tech Center for Quantum Information Science and Engineering, Blacksburg, VA 24061, USA}

\author{Sophia E. Economou\orcidlink{0000-0002-1939-5589}}
\email{economou@vt.edu}
\affiliation{Department of Physics, Virginia Tech, Blacksburg, VA 24061, USA}
\affiliation{Virginia Tech Center for Quantum Information Science and Engineering, Blacksburg, VA 24061, USA}

\date{\today}% It is always \today, today,

\begin{abstract}
Preparing Gibbs states, which describe systems in equilibrium at finite temperature, is of great importance, particularly at low temperatures. In this work, we propose a new method-- TEPID-ADAPT-- that prepares the thermal Gibbs state of a given Hamiltonian at low temperatures using a variational method that is partially adaptive and uses a purification with a minimal number of ancillary qubits. We also present an alternative implementation without ancillary qubits. A key technical innovation here is to use a mixed-state ansatz where the entropy can be efficiently calculated, with no computational overhead. Our algorithm uses a truncated, parametrized eigenspectrum of the Hamiltonian. Beyond preparing Gibbs states, this approach also straightforwardly gives us access to the truncated low-energy eigenspectrum of the Hamiltonian, making it also a method that prepares excited states simultaneously. As a result of this, we are also able to prepare thermal states at any lower temperature of the same Hamiltonian without further optimization.
\end{abstract}

\maketitle

%%%%%%%%%%%%%%%%%%%%%%%%%%%%%%%%%%%%%%%%%%%%%%%%%%%%%%%%%%%%%%%%%%%%%%%%%%%%%%%%%%%%%%%%%%
%%%%%%%%%%%%%%%%%%%%%%%%%%%%%%%%%%%%%%%%%%%%%%%%%%%%%%%%%%%%%%%%%%%%%%%%%%%%%%%%%%%%%%%%%%

\section{Introduction}\label{sec:Intro}
The problem of state preparation is a crucial subroutine in several quantum algorithms, both near-term and fault-tolerant. Therefore, finding efficient ways to prepare quantum states is of great interest. The majority of currently available techniques focus on the preparation of pure states of closed systems, most of which tackle ground-state preparation. In a more realistic setting, quantum systems inevitably have a coupling to an environment. This makes the effective state of the system of interest mixed, described by a density matrix. As a result, addressing the problem of density matrix preparation is an important milestone in the simulation of real physical systems using quantum computers. Of particular interest are thermal Gibbs states, which describe the state of an open quantum system at thermal equilibrium with a bath held at finite temperature

\begin{equation}
    \rho_G\equiv \frac{e^{-\beta\,{H}}}{Z},\qquad Z=\Tr(e^{-\beta\,{H}}),
\end{equation}
where $H$ is the Hamiltonian of the system of interest, and $\beta$ is the inverse temperature. The Gibbs state uniquely minimizes the free energy

\begin{equation}
    F=\langle {H}\rangle-\beta^{-1}\,S,
    \label{eq:FEnergy}
\end{equation}
where $\langle{\,\boldsymbol{\cdot}\,}\rangle\equiv\Tr(\rho\,\boldsymbol{\cdot})$, and $S=-\Tr(\rho\log\rho)$ is the von Neumann entropy. Preparing Gibbs states is important for applications such as quantum simulation~\cite{ChildsPNAS2018}, quantum chaos~\cite{Garcia-Mata2023}, combinatorial optimization problems~\cite{KirkpatrickScience1983,SommaPRL2008}, and quantum machine learning~\cite{KieferovaPRA2017, BiamonteNature2017}. From the perspective of state preparation, the low-temperature regime is more interesting and challenging~\cite{Watrous2008,AharanovACM2013,BakshiIEEE2024}. In this regime, only the low-energy eigenstates of the Hamiltonian contribute significantly to the Gibbs state. Preparing excited states, particularly the low-lying eigenstates that make up the low-temperature Gibbs states of a Hamiltonian is a closely related task of great importance for applications in both physics and chemistry. It has applications including understanding dynamics of systems, measuring transition and decay rates~\cite{IbePhysRevR2022, CiavarellaPRD2020}, and mass gaps of quantum field theories~\cite{Chandani2024}. Existing methods include quantum phase estimation (QPE)~\cite{KitaevECCC1995}, variational quantum algorithms~\cite{HiggottQuantum2019, NakanishiPhysRevR2019,WenQuantEng2021, XieJCTC2022, Sherbert2022,GochoNPJCompMat2023,Xu:2023dgi,GrimsleyQST2025,Chandani2024}, and subspace methods~\cite{McCleanPRA2017, MottaNaturePhys2019, Parrish2019, StairJCTC2020, AsthanaRSCCS2023, CianciJCTC2024}. Ref.~\cite{Xu:2023dgi} uses a purification scheme to prepare an equal mixture of states to target low-lying states simultaneously.

Existing methods to study Gibbs states on quantum computers include Lindblad simulation and sampling~\cite{PoulinPRL2009,TemmeNature2011,ChowdhuryQIC2017, Chen2023_1, Chen2023_2,EassaNPJQI2024, BergamaschiIEEEFOCSProceedings2024, Rajakumar2024,Brunner2024}, imaginary time evolution~\cite{MottaNaturePhys2019, SunPRXQ2021, KamakariPRXQ2022,GetelinaSciPostPhys2023}, and state purification. Out of these, state purification is the most intuitive and amenable to near-term applications. The broad idea is to purify the mixed state of the effective open system by enlarging the Hilbert space using ancillary qubits. Unitary operations can then be used to find a pure state on the extended system that prepares the Gibbs state on the system register, after tracing out the ancillary qubits. Known variational methods can be used to find the purified state. A particular purification, the thermofield double (TFD) state, is interesting in its own right from the perspective of holography~\cite{MaldacenaJHEP2003, MaldacenaFdP2013}. Preparing the TFD state, by definition, requires doubling the number of system qubits for purification~\cite{CottrellJHEP2019,WuPRL2019,SagastizabalNPJQI2021}. Variationally preparing the Gibbs state this way can have the advantage of mapping to a problem of finding the ground state of a Hamiltonian, albeit in a larger Hilbert space~\cite{CottrellJHEP2019}. However, the cost of doubling the system size is excessive, especially if the goal is to prepare Gibbs states at low-temperatures. Moreover, the extended Hamiltonian involves heuristic assumptions about the interaction between the two halves of the expanded system, and potentially leads to a much more complicated Hamiltonian.

\begin{figure*}[t!]
    % \centering
    \includegraphics[width=\linewidth]{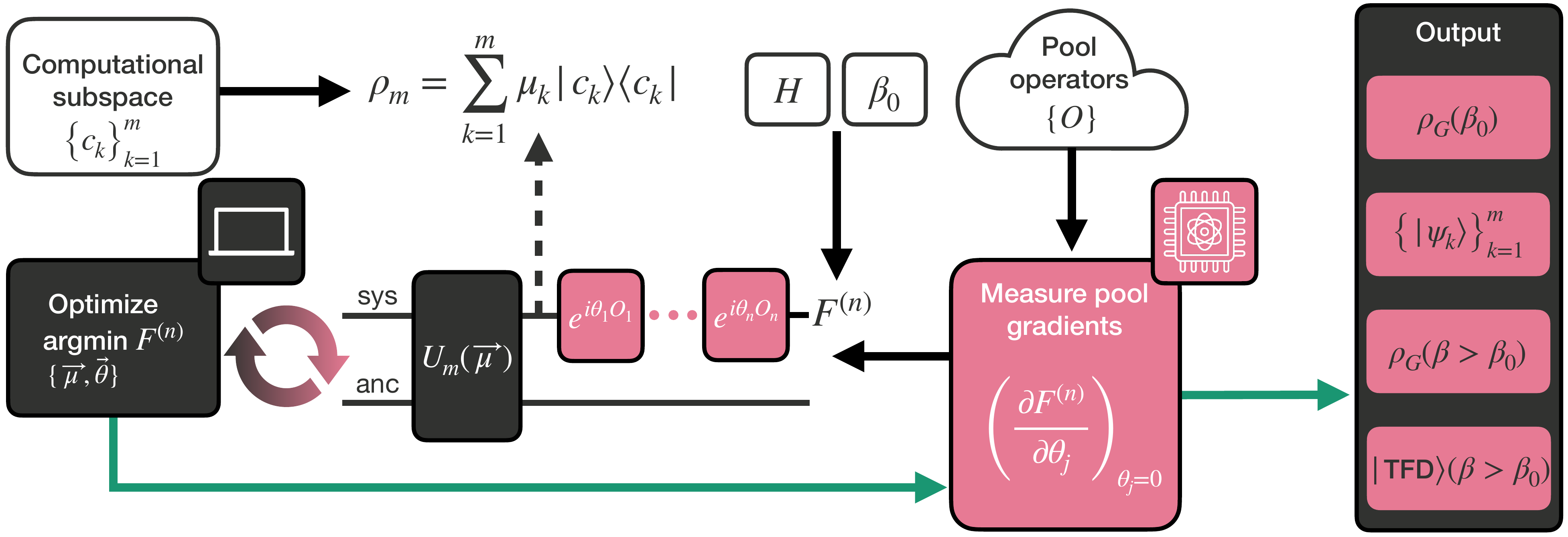}
    \caption{A diagrammatic workflow for TEPID-ADAPT. The white shapes indicate the inputs-- the Hamiltonian $H$, the inverse temperature $\beta_0$, an $m$-dimensional computational subspace, and an operator pool for ADAPT. The black circuit, labeled $U_m(\vec{\mu})$ is the static part of the ansatz, which prepares $\rho_m$ on the system register. The pink circuit parametrized by $\vec{\theta}$ is the adaptively generated portion that maps the computational subspace to the truncated eigenspace of the Hamiltonian. The two solid green arrows (from left-right) respectively indicate the convergence of the VQE subroutine and the pool gradients vanishing. The primary output of TEPID-ADAPT is the target Gibbs state at $\rho_G(\beta_0)$. We can also readily prepare the low-lying eigenstates of the Hamiltonian in the truncated subspace $\{\ket{\psi_k}\}_{k=1}^m$, the Gibbs states at lower temperatures $\rho_G(\beta>\beta_0)$, and the TFD state at inverse temperatures $\beta\geq\beta_0$.}
    \label{fig:SummaryChart}
\end{figure*}

Variational methods that prepare Gibbs states using other purifications face an obstacle at the level of the cost function. The free energy Eq.~\eqref{eq:FEnergy} is difficult to measure even on quantum hardware because measuring the von Neumann entropy of an arbitrary density matrix requires at least partial state tomography. To get around this difficulty, Ref.~\cite{WangPRApp2021} uses a truncated  Taylor series of the free energy as the cost function, while Ref.~\cite{Warren2022} uses a different cost function that measures proximity to a truncated version of the target Gibbs state. In Ref.~\cite{ConsiglioPRA2024}, the free energy \textit{without} any truncation is used as a cost function. This is enabled by a clever modular construction of the parametrized quantum circuit that first prepares
a diagonal mixed state on the system qubits, followed
by operations that leave the von Neumann entropy invariant. However, the use of a hardware-efficient ansatz for the first block of the circuit and the fact that more ancilla qubits than necessary to prepare an accurate Gibbs state at low-temperatures are used introduce ad hoc elements to the algorithm, additional measurements, and a classical post processing overhead. In Ref.~\cite{GrimsleyQST2025}, multiple eigenstates of chemical systems are targeted using an ensemble of orthogonal reference states, together with a weighted sum of their evolved energies as a cost function. However, a classical Ritz diagonalization~\cite{HendekovicChemPhysLett1982} is needed to diagonalize the Hamiltonian in the subspace, because the variationally trained unitary does not yield eigenstates directly, complicating the task of preparing low-energy eigenstates on the quantum computer. Moreover, the weights of the states are fixed beforehand and remain static throughout the optimization, making this method unsuitable for preparing Gibbs states, unless perhaps the energy eigenvalues are already known via some other means.

In this work, we introduce an algorithm that addresses all these issues simultaneously. Our method uses minimal ancillary qubits to prepare Gibbs states at low temperatures as well as the low-energy eigenstates that the Gibbs state is primarily composed of. We propose a partially adaptive variational quantum method that uses a parametrized truncation of the Hamiltonian’s spectrum. While the broad modular structure of our ansatz is the same as the one used in Ref.~\cite{ConsiglioPRA2024}, we use a parametrization in the first (static) part of our ansatz that gives us the von Neumann entropy without the need for classical post-processing. This parametrization also gives us a systematic way to reduce the number of ancillas needed to prepare Gibbs states in the low-temperature regime. Once the parametrized mixed state that is diagonal in the computational basis is prepared, we variationally find a unitary with support purely on the system qubits that transforms the state into the truncated eigenbasis of the Hamiltonian. While one could in principle use a static ansatz to do this, we leverage the power of ADAPT-VQE~\cite{GrimsleyNatureComm2019}, given its ability to find shorter-depth circuits. This readily gives us access to the low-lying eigenspectrum of the Hamiltonian, upon successful preparation of the target Gibbs state. This feature also facilitates the preparation of Gibbs states at lower temperatures \textit{without any further parameter optimization}. Moreover, we can prepare low-temperature TFD states without any further optimization. While most existing approaches use the TFD state as a means to prepare the Gibbs state, thereby explicitly requiring the doubling of the system qubits, our method prepares a low-temperature Gibbs state using a minimal ancillary overhead, and uses the corresponding unitaries to readily prepare the TFD state. To our knowledge this is the first variational quantum algorithm to prepare the TFD state that does not necessitate variational optimization on the doubled system. We also propose an alternative, ancilla-free implementation of our algorithm, which uses the interpretation of a mixed state as a convex sum of pure states. In this approach, computational basis states are sampled from a parametrized probability distribution. The expectation value of the Hamiltonian, which is needed for the cost function (given by the Gibbs free energy), is measured as an ensemble average over the samples. This interpretation of mixed states has been studied in other contexts in Refs.~\cite{EzzellQST2023,ChenPRA2025}. In the near-term, this implementation of our algorithm would involve shallower circuits. This approach is related to Ref.~\cite{GrimsleyQST2025}, but in comparison has two key advantages: the freedom in TEPID-ADAPT to variationally optimize the weights of individual states, which is necessary for Gibbs state preparation, and not incurring an additional classical overhead associated with diagonalizing the Hamiltonian in the subspace.

The organization of the paper is as follows. In Sec.~\ref{sec:VarMethods}, we briefly review variational quantum algorithms, with an emphasis on adaptive methods. In Sec.~\ref{sec:TEPID}, we introduce our method TEPID-ADAPT and highlight some of its key features. In Sec.~\ref{sec:Results}, we apply our method to different phases of a standard spin system---the 1D quantum Heisenberg XXZ model---and compare the performance against exact diagonalization. We then conclude in Sec.~\ref{sec:Conclusions}, and discuss further directions to pursue.

%%%%%%%%%%%%%%%%%%%%%%%%%%%%%%%%%%%%%%%%%%%%%%%%%%%%%%%%%%%%%%%%%%%%%%%%%%%%%%%%%%%%%%%%%%
%%%%%%%%%%%%%%%%%%%%%%%%%%%%%%%%%%%%%%%%%%%%%%%%%%%%%%%%%%%%%%%%%%%%%%%%%%%%%%%%%%%%%%%%%%

\section{Variational quantum algorithms}\label{sec:VarMethods}
VQAs have four main ingredients: a parametrized quantum circuit (we will refer to this as the ansatz going forward), a cost function that can be measured on a quantum computer, a reference state, and a classical optimizer that informs the variation of the parameters in the ansatz. The classical optimizer could be gradient-based or gradient-free. VQAs, being a class of hybrid algorithms, leverage both quantum and classical resources. Quantum computers are used to efficiently evolve and measure observables such as the cost function and its gradients. These are fed to the classical optimizer, which suggests updates to the parameters in the ansatz to lower the cost function. For preparing pure states, starting with a computational basis state, for instance $\ket{0\cdots 0}\bra{0\cdots 0}$, is a valid choice because all pure states that are drawn from the same Hilbert space are \textit{unitarily equivalent} to one another, which we denote compactly as
\begin{equation}
    \ket{\psi_1}\bra{\psi_1}\Ueq\ket{\psi_2}\bra{\psi_2}.
\end{equation}
Note that this is a statement about the existence of a unitary, and does not account for the computational cost of implementing it. VQAs aim to solve the optimization problem
\begin{equation}\label{eq:CFunctionVQA}
    \vec{\theta}_c=\argminA_{\theta}\mathcal{C}(\vec{\theta})\equiv\Tr\left({\mathcal{O}}\,V(\vec{\theta})\ket{\psi_0}\bra{\psi_0}V^{\dagger}(\vec{\theta})\right),
\end{equation}
where $\ket{\psi_0}\bra{\psi_0}$ is the initial reference state, and ${\mathcal{O}}$ is the operator that is measured to yield the cost function.

The choice of ansatz is an important ingredient in VQAs. A common approach is to use a static ansatz~\cite{PeruzzoNature2014, WeckerPRA2015, KandalaNature2017,GardNPJ2020, BurtonPRR2024}, where a predefined sequence of parameterized gates  are optimized until the convergence criteria are met.

%%%%%%%%%%%%%%%%%%%%%%%%%%%%%%%%%%%%%%%%%%%%%%%%%%%%%%%%%%%%%%%%%%%%%%%%%%%%%%%%%%%%%%%%%%

\subsection{ADAPT-VQE}

Alternatively, one could use an ansatz that is adaptively generated~\cite{GrimsleyNatureComm2019, GrimsleyNPJQI2022} using a pool of predefined operators. The operator pool can be tailored for the problem at hand. For instance, a pool could consist of operators that respect some symmetry of the Hamiltonian, or could be comprised of operators related to dominant interactions in the cost Hamiltonian. The key feature of adaptive VQAs is that the ansatz is not predetermined; it is created dynamically by adding operators one (or a few~\cite{AnastasiouPRR2024}) at a time, each one being  chosen based on the local gradients of the pool operators in the cost function landscape. Measuring these gradients involves some parallelizable overhead on quantum processors. All the parameters are re-optimized after operators are added. This process is repeated until preset convergence criteria are met. This has been shown to be very effective for state preparation for problems in chemistry and physics. Adaptive methods are shown to yield shorter depth circuits~\cite{TangPRXQ2021,FeniouSpringerCP2023,AnastasiouPRR2024,Ramoa2024}. They are also resistant to issues associated with difficult optimization landscapes~\cite{GrimsleyNPJQI2022,AnastasiouPRR2024,SherbertIEEEConferenceProceeding2024}. 

%%%%%%%%%%%%%%%%%%%%%%%%%%%%%%%%%%%%%%%%%%%%%%%%%%%%%%%%%%%%%%%%%%%%%%%%%%%%%%%%%%%%%%%%%%

\subsection{Variational optimization of mixed states}

When it comes to variationally preparing mixed states, one of the primary challenges is finding the correct unitary equivalence class of the target state $\rho$. If the unitary equivalence class is known apriori, one could start with a reference state, $\rho_0$, within it and variationally optimize a unitary, $V$, that minimizes an appropriate cost function, similar to the case of pure states: 
\begin{equation}\rho_0\Ueq\rho\iff\exists\,V:V\rho_0V^{\dagger}=\rho.
\end{equation}
However, knowing the unitary equivalence class requires knowledge of the eigenvalues of the target density matrix, of which there are an exponentially large number.

This is typically addressed by state purification using ancillary qubits and ans\"{a}tze that span the extended system~\cite{CottrellJHEP2019,WuPRL2019,WangPRApp2021,SagastizabalNPJQI2021,Warren2022,ConsiglioPRA2024}. This allows the system density matrix to explore different unitary equivalence classes until the correct one is found. 

Another important problem is the choice of an appropriate cost function. For the Gibbs state, the free energy (Eq.~\eqref{eq:FEnergy}) is a natural choice, as it admits a variational principle since the Gibbs state uniquely minimizes it. The roadblock then, is that the free energy is not a quantum mechanical observable due to the von Neumann entropy term in it. Our algorithm provides a constructive way to get around this.

%%%%%%%%%%%%%%%%%%%%%%%%%%%%%%%%%%%%%%%%%%%%%%%%%%%%%%%%%%%%%%%%%%%%%%%%%%%%%%%%%%%%%%%%%%
%%%%%%%%%%%%%%%%%%%%%%%%%%%%%%%%%%%%%%%%%%%%%%%%%%%%%%%%%%%%%%%%%%%%%%%%%%%%%%%%%%%%%%%%%%

\section{TEPID-ADAPT}\label{sec:TEPID}

In this section, we introduce our partially adaptive method for variationally preparing low-temperature thermal Gibbs states of a given Hamiltonian. We call our method TEPID-ADAPT, which stands for \textbf{T}runcated \textbf{E}igenvalue \textbf{P}arametrized \textbf{I}nitial \textbf{D}ensity. In Fig.~\ref{fig:SummaryChart}, we show a diagrammatic workflow of our algorithm, that highlights its various key features.

At low temperatures (large $\beta$), only the low-lying energy eigenstates of the Hamiltonian contribute significantly to the Gibbs state. As a result, an approximation of the Gibbs state is
\begin{equation}\label{eq:GibbsTrunc}
    \rho_G\equiv \frac{e^{-\beta\,H}}{Z}\approx \frac{1}{Z_m}\sum_{k=1}^{m}e^{-\beta\,E_k}\ket{\psi_k}\bra{\psi_k},
\end{equation}
where $\left\{E_k,\ket{\psi_k}\right\}$ is the ordered eigenspectrum of the Hamiltonian $H$, $m$ is the truncation of the eigenspectrum, and $Z_m$ is the truncated partition function. In this work, we treat $m$, which sets an upper bound on the amount of entanglement between the system and ancillary qubits, as a hyperparameter:
\begin{equation}
    S\leq\log m.
\end{equation}
Consider the diagonal state in the computational basis
\begin{equation}
    \rho_0=\frac{1}{Z_m}\sum_{k=1}^me^{-\beta\,E_k}\ket{c_k}\bra{c_k},
\end{equation}
where $\left\{\ket{c_k}\right\}_{k=1}^m$ is a set of computational basis elements. We note that $\rho_0$ is approximately unitarily equivalent to the Gibbs state
\begin{equation}
    \rho_0\Uapprox\rho_G.
\end{equation}
This unitary equivalence would be exact if $m=2^{N_s}$, where $N_s$ is the number of system qubits. Inspired by this, we construct a parameterized reference state on the system register that is diagonal in the computational basis with $m$ non-zero eigenvalues:
\begin{equation}\label{eq:rho_m}
    \rho_m\equiv\sum_{k=1}^m\mu_k\ket{c_k}\bra{c_k},
\end{equation}
where the parameters $\{\mu_k\}_{k=1}^m$ are variationally optimized over our algorithm, subject to the constraint $\sum_{k=1}^m \mu_k=1$. With this parametrized $\rho_m$ as the reference state, TEPID-ADAPT aims to do two things:
\begin{enumerate}
    \item Find the correct approximate unitary equivalence class of the target Gibbs state:
    \begin{equation}
        \mu_k\to\frac{e^{-\beta E_k}}{Z_m}\quad\forall\, k=1,\cdots,m.
    \end{equation}
    \item Find a unitary operation on the system register that rotates the computational basis to the truncated eigenbasis of $H$:
    \begin{equation}
        V_A:\ket{c_k}\to\ket{\psi_k}\quad\forall\,k=1,\cdots,m.
    \end{equation}
\end{enumerate}

\begin{figure}[ht!]
    \centering
    \begin{quantikz}[classical gap=0.07cm]
        \lstick{$\ket{0}^{\otimes N_s}$} & \qwbundle{N_s} & \gate[2]{U_m(\vec{\mu})}&\slice{prep} & & \gate{V_A(\vec{\theta})}&\rstick{$\approx\rho_G$}\\
        \lstick{$\ket{0}^{\otimes N_a}$} & \qwbundle{N_a} & & & & & \trash{\text{trace}}
    \end{quantikz}
    \caption{General block diagram for the variational ansatz of TEPID-ADAPT. $U_m(\mu)$ prepares $\rho_m$ on the system register (top), as indicated by the red line. $V_A(\vec{\theta})$ is an adaptively generated unitary on the system register that approximately evolves $\rho_m$ to the target Gibbs state.}
    \label{fig:BlockAnsatz}
\end{figure}
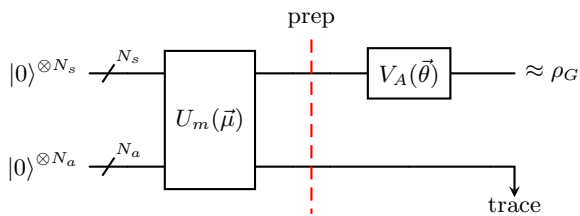

We achieve these using a blocked ansatz shown in Fig~\ref{fig:BlockAnsatz}. The first block $U_m(\vec{\mu})$ prepares the parametrized density matrix $\rho_m$ in Eq.~\eqref{eq:rho_m} on the system register. We will refer to $U_m(\vec{\mu})$ as the initial state-preparation unitary. The rest of the ansatz is adaptively generated with support purely on the system register. We denote this block by $V_A(\vec{\theta})$ (the subscript stands for adaptive), where $\vec{\theta}$ are the variational parameters of this unitary. This separation and nomenclature fit well within the language of VQA's---$U_m(\vec{\mu})$ prepares a reference state, and $V_A(\vec{\theta})$ rotates it. The key difference is that the reference state itself is parametrized and optimized over. This degree of freedom is necessary since the unitary equivalence class of the target Gibbs state is unknown apriori.

While we build $V_A(\vec{\theta})$ adaptively in this work, note that one could instead use a static ansatz. Upon convergence, this becomes a unitary that approximately rotates $\rho_m$ to $\rho_G$ in Eq.~\eqref{eq:GibbsTrunc}. The cost function we use for this VQA is the free energy in Eq.~\eqref{eq:FEnergy}. Note that this is not of the form written in Eq.~\eqref{eq:CFunctionVQA} because the von Neumann entropy is not a quantum observable. Therefore, efficiently estimating it is a difficult problem. In our case, this is made easy by construction of the ansatz: Since $\rho_m(\vec{\mu})$ is diagonal in the computational basis, the von Neumann entropy is simply the Shannon entropy: 
\begin{equation}
    S=-\sum_{k=1}^m\mu_k\log\mu_k.
\end{equation}
Moreover, in an instance of the ansatz, $S$ remains unchanged beyond the dashed vertical line in Fig.~\ref{fig:BlockAnsatz}. This is because the $V_A(\vec{\theta})$ part of the ansatz only has support on the system register, and the von Neumann entropy of a density matrix is invariant under unitary transformations. Because of this construction, the energy $\langle H\rangle$ is the only piece of the cost function that needs to be measured on the quantum computer, where the cost function is the free energy:
\begin{equation}\label{eq:FEnergtparam}
    F(\vec{\mu},\vec{\theta})=\langle H\rangle(\vec{\mu},\vec{\theta})-\beta^{-1}S(\vec{\mu}).
\end{equation}
This makes the measurement complexity in TEPID-ADAPT identical to that of a regular VQE for ground-state preparation. We emphasize here that after the addition of an operator to the ansatz, we re-optimize \textit{all} the parameters in the circuit, including the $\vec{\mu}$ parameters. In other words, at each optimization step, both the unitary equivalence class of $\rho_m$ and the adaptive unitary $V_A$ are allowed to change. In Appendix~\ref{app:Theorem1}, we prove that the rank-$m$ state that minimizes the free energy is the mixture of the lowest $m$ eigenstates of $H$ weighted by the Boltzmann factors, as shown in Eq.~\eqref{eq:GibbsTrunc}. 

The choice of computational basis elements $\{c_k\}_{k=1}^m$ that go into $\rho_m$ is essentially a choice on the form of the static state-preparation unitary $U_m(\vec{\mu})$. This choice does not affect the unitary equivalence class for a given set of $\vec{\mu}$ parameters. While any choice of $\{c_k\}_{k=1}^m$ should work in principle, provided the operator pool for $V_A$ is expressive enough, a careful choice of computational basis elements has an impact on the complexity of $V_A$. In Sec.~\ref{sec:Results}, we use knowledge of the phase diagram of the Heisenberg XXZ model to choose the computational basis elements.

The construction of $U_m$ depends on what family of $\vec\mu$ we consider. In Appendix~\ref{app:ExplicitUm} we describe how to implement $U_m$   for an arbitrary probability distribution $\vec\mu$ over $[m]$. Analytic gradients of the cost function can be computed with effort similar to that in the standard VQE, as we describe in Appendix~\ref{App:AnalyticalGradients}. Prior to the addition of operators adaptively to the basis change unitary $V_A$, there is a choice to be made on what the initial set of parameters $\vec{\mu}$ for the state preparation unitary $U_m(\vec{\mu})$ is. This choice is addressed in Appendix~\ref{App:InitialState}.

We also provide an alternative, ancilla-free implementation of TEPID-ADAPT in Appendix~\ref{app:AncillaFree} that uses the interpretation of a mixed state as a convex sum of pure states, and measurements as ensemble averages. This approach is more resource-efficient both in number of qubits, and in gate depth, since the circuit that prepares $\rho_m$ is replaced with a classical sampling overhead, and preparation of the sampled computational basis state. The unitary $V_A(\vec{\theta})$ is trained via measurement of the cost function and its gradients as ensemble averages. Once $V_A(\vec{\theta})$ is trained, low-temperature thermal observables can be measured readily by weighting the corresponding measurements on the individual states by the Boltzmann factors. The key difference between these approaches is that the ancilla-free version prepares the Gibbs state as a mixed state instead of preparing its purification. While the mixed state $\rho_m$ is cheaper to prepare than the purification, it also can be weaker in some settings. For example, preparing the purification can be used for faster gradient estimation~\cite{Bittel2022}, by replacing the sampling over the Gibbs state by coherent amplitude amplification. Moreover, in some applications, our goal might be to prepare a specific purification, such as the TFD (Thermofield Double), which we discuss further in Sec.~\ref{subsec:TFD}. It is also possible to train using the ancilla-free algorithm and then, if it is needed, use ancillas to prepare the TFD state for an application. Which approach is best will depend on the larger context within which our algorithm is used.

%%%%%%%%%%%%%%%%%%%%%%%%%%%%%%%%%%%%%%%%%%%%%%%%%%%%%%%%%%%%%%%%%%%%%%%%%%%%%%%%%%%%%%%%%%

\subsection{Key features}\label{subsec:Features}

In this section, we highlight two important features of TEPID-ADAPT. Let us denote the adaptive portion of the \textit{converged} unitary operation by $V_A(\vec{\theta}^*)$ for a given inverse temperature $\beta_0$. Recall that $V_A^{\dagger}(\vec{\theta}^*)$ diagonalizes $H$ in the $m$-truncated subspace. As shown in Fig~\ref{fig:BlockAnsatz}, this unitary acts purely on the system register. The first feature we expand on is how we obtain access to the excited states that contribute significantly to the Gibbs state. To prepare these eigenstates $\{\ket{\psi_k}\}$, we simply evolve computational basis states by the converged unitary
\begin{equation}\label{eq:EigPrep}
    V_A(\vec{\theta}^*)\ket{c_k}\equiv\ket{\psi_k}\,\quad\forall\,\,k=1,\cdots,m,
\end{equation}
where $\{\ket{c_k}\}_{k=1}^m$ is the set of computational basis states that we started with in Eq.~\eqref{eq:rho_m}. To find the energies, we can measure the expectation values of the Hamiltonian on the prepared eigenstates:
\begin{equation}\label{eq:EnergyAccess}
    E_k = \bra{\psi_k}H\ket{\psi_k}\,\quad\forall\,\,k=1,\cdots,m.
\end{equation}
Alternatively, we can obtain all the energy differences $\Delta E_k=E_k-E_0$ via the ratios of the $\vec{\mu}$ parameters. So, we only \textit{need} to measure the ground state energy using Eq.~\eqref{eq:EnergyAccess} to obtain the low-lying energies.

The next feature we highlight is the ability to prepare the Gibbs state at any lower temperature with a fidelity that increases with $\beta$. This feature is a direct consequence of having access to the truncated eigenspectrum. Intuitively, this provides us with the relevant information regarding the interpolation between the Gibbs state at $\beta_0$ and the Gibbs state at $\beta\to\infty$, corresponding to just the ground state. More concretely, we note two things:
\begin{enumerate}
    \item Excited states become increasingly less important for the Gibbs state as we lower the temperature.
    \item $V_A^{\dagger}(\vec{\theta}^*)$ diagonalizes any function of $H$ in the $m$-truncated subspace.
\end{enumerate}
As a result, $V_A(\vec{\theta}^*)$ can be used without any parameter re-optimization to prepare the Gibbs state at any lower temperature $\beta>\beta_0$, provided we prepare the corresponding $\rho_m$ first using $U_m(\vec{\mu}')$, where
\begin{align}\label{eq:ModifiedStatePrepParams}
    \mu'_k&=\frac{1}{Z_m}e^{-\beta E_k}\,\quad\forall\,\,k=1,\cdots,m\\
        &=\frac{1}{1+\sum_{j=2}^me^{-\beta \Delta E_j}}e^{-\beta \Delta E_j},
\end{align}
where the energy differences $
\Delta E_j=E_j-E_1$ are found using ratios of the $\vec{\mu}$ parameters:
\begin{equation}
    \Delta E_k = \frac{1}{\beta}\log\frac{\mu_1}{\mu_k}\,\quad\forall\,\,k=1,\cdots,m.
\end{equation}
Moreover, the fidelity with the corresponding Gibbs state improves as we increase $\beta$ because of point 1 above.

%%%%%%%%%%%%%%%%%%%%%%%%%%%%%%%%%%%%%%%%%%%%%%%%%%%%%%%%%%%%%%%%%%%%%%%%%%%%%%%%%%%%%%%%%%

\subsection{Preparing TFD states}\label{subsec:TFD}

In this section, we show how to prepare low-temperature thermofield double (TFD) states using TEPID-ADAPT without further variational optimization. The TFD state of an $N_s$-qubit Hamiltonian $H$ is a pure state that lives on $2N_s$ qubits, and is defined as 
\begin{equation}
    \ket{\text{TFD}}\equiv\frac{1}{\sqrt{Z}}\sum_{k=1}^{2^{N_s}}e^{-\beta\,E_k/2}\ket{\psi_k}_L\otimes\ket{\overline{\psi}_k}_R,
\end{equation}
where $\{E_k\}_{k=1}^{2^{N_s}}$ are the eigenenergies of $H$, and $\{\ket{\psi_k}\}_{k=1}^{2^{N_s}}$, $\{\ket{\overline{\psi}_k}\}_{k=1}^{2^{N_s}}$ are respectively the eigenstates of $H$ and $\overline{H}$, with $\overline{\,\,\cdot\,\,}$ representing complex conjugation. The truncated eigenspace of $\overline{H}$ can be accessed using the complex conjugate of the converged basis-change unitary $V_A(\vec{\theta})$ used to prepare the low-temperature Gibbs state of $H$: 
\begin{equation}
    \overline{V}_A(\vec{\theta}^*)\ket{c_k}\equiv\ket{\overline{\psi}_k}\quad\forall k=1,\cdots,m.
\end{equation}
$\overline{V}_A(\vec{\theta}^*)$ can be readily constructed by reversing the sign of the $\vec{\theta}^*$ parameters in front of generators that have an even number of $\sigma^y$ Pauli operators.
Upon tracing out either half ($L$ or $R$) of the doubled system on the TFD state, one obtains the Gibbs state at inverse temperature $\beta$ on the other half:
\begin{equation}
    \Tr_{L,R}\ket{\text{TFD}}=\rho_G^{R,L}.
\end{equation}
Going forward, we will drop the $L,R$ indices for convenience. In the low-temperature (large $\beta$) regime, a good approximation of the $\ket{\text{TFD}}$ state is
\begin{equation}\label{eq:TFDTrunc}
    \ket{\text{TFD}}\approx\frac{1}{\sqrt{Z_m}}\sum_{k=1}^m e^{-\beta\,E_k/2}\ket{\psi_k}\otimes\ket{\overline{\psi}_k},
\end{equation}
where we have truncated the sum to only contain the $m$ lowest eigenstates of $H$ and $\overline{H}$, and $Z_m$ is the truncated partition function. We can readily prepare this state using TEPID-ADAPT using the same converged unitaries $U_m(\vec{\mu}^*)$ and $V_A(\vec{\theta}^*)$ from preparing the Gibbs state at inverse temperature $\beta$, with
\begin{equation}
    \mu_k^*\equiv\frac{1}{Z_m}e^{-\beta\,E_k}.
\end{equation}

The circuit for preparing the approximate TFD state in Eq.~\eqref{eq:TFDTrunc} is shown in Fig.~\ref{fig:TFDCircuit}. We begin by introducing a second ancillary register with $N_s-N_a$ qubits, all initialized in the $\ket{0}$ state. The state of the the full system is
\begin{equation}
    \ket{\Omega}\equiv\frac{1}{\sqrt{Z_m}}\sum_{k=1}^me^{-\beta\,E_k/2}\ket{c_k}\otimes\ket{k-1}\otimes\ket{0}^{\otimes(N_s-N_a)}.
\end{equation}
Next, we apply a permutation matrix to the ancillary registers. These are a
class of matrices that permute the computational basis. The particular permutation matrix $P ({\ket{c_k}})$ we need is the one that maps the first $m$ computational basis elements $\{\ket{k-1}\otimes\ket{0}^{\otimes(N_s-N_a)}\}_{k=1}^m$ to the chosen computational subspace $\{\ket{c_k}\}_{k=1}^m$, giving us the state
\begin{equation}
    \ket{\Phi}=\frac{1}{\sqrt{Z_m}}\sum_{k=1}^m e^{-\beta\,E_k/2}\ket{c_k}\otimes\ket{c_k}.
\end{equation}
Finally, we act the converged unitary $V_A(\vec{\theta}^*)$ and $\overline{V}_A(\vec{\theta}^*)$ respectively on the $L$ and $R$  registers of $N_s$ qubits to obtain the approximate low-temperature TFD state as in Eq.~\eqref{eq:TFDTrunc}. The key advantage of preparing the TFD state this way is the avoidance of doing VQE on the doubled system, making it a very valuable technique in the low-temperature regime. 

\begin{figure}[ht!]
    \centering
    \begin{quantikz}[classical gap=0.07cm]
        \lstick{$\ket{0}$} & \qwbundle{N_s} & \gate[2]{U_m(\vec{\mu}^*)} & & & \gate{V_A(\vec{\theta}^*)} & \rstick[3]{\rotatebox{90}{$\ket{\text{TFD}}$}}\\
        \lstick{$\ket{0}$} & \qwbundle{N_a} & & \gate[2]{P\left(\left\{\ket{c_k}\right\}\right)}  \\
        \lstick{$\ket{0}$} & \qwbundle{N_s-N_a} & & & \qwbundle{N_s} & \gate{\overline{V}_A(\vec{\theta}^*)} &
    \end{quantikz}
    \caption{The circuit for preparing a low-temperature TFD state using TEPID-ADAPT without any further variational optimization. The parameters in the unitaries are the converged ones from preparing the low-temperature Gibbs state. $P({\{\ket{c_k}}\})$ is a permutation matrix that maps the first $m$ computational basis states on the ancillary registers to the chosen elements $\{\ket{c_k}\}_{k=1}^m$.}
    \label{fig:TFDCircuit}
\end{figure}

%%%%%%%%%%%%%%%%%%%%%%%%%%%%%%%%%%%%%%%%%%%%%%%%%%%%%%%%%%%%%%%%%%%%%%%%%%%%%%%%%%%%%%%%%%
%%%%%%%%%%%%%%%%%%%%%%%%%%%%%%%%%%%%%%%%%%%%%%%%%%%%%%%%%%%%%%%%%%%%%%%%%%%%%%%%%%%%%%%%%%

\section{Results}\label{sec:Results}

In this section, we showcase our method TEPID-ADAPT, along with its key features. The results presented here are obtained using noiseless classical simulations. We use the Heisenberg XXZ model with open boundary conditions given by the Hamiltonian
\begin{equation}
    H_{\text{XXZ}}=\sum_{k=1}^{N_s-1}\sigma_{k}^x\sigma_{k+1}^x+\sum_{k=1}^{N_s-1}\sigma_{k}^y\sigma_{k+1}^y+J_z\sum_{k=1}^{N_s-1}\sigma_{k}^z\sigma_{k+1}^z,
\end{equation}
where $J_z$ is the nearest-neighbor coupling strength of the $ZZ$ interaction, and $\sigma^{x,y,z}$ are the Pauli matrices. This model has ferromagnetic, paramagnetic, and antiferromagnetic phases. This is an integrable model~\cite{BetheZeitschriftPhysik1931} and serves as a good test ground for our method. We will present results for parameters corresponding to the three phases. The operator pool we use here to generate the adaptive part of the ansatz, $V_A(\vec{\theta})$, is the full Pauli pool, which is the set of all $N_s$ qubit Pauli operators, which has size $4^{N_S}$. This pool is not scalable to large systems but maximizes the flexibility in the adaptive protocol, so it is appropriate for this proof of concept. In the results below, we consider a system with $N_s=6$, and $\beta_0=3.0$.

\begin{figure}[ht!]
    \centering
    \includegraphics[width=\linewidth]{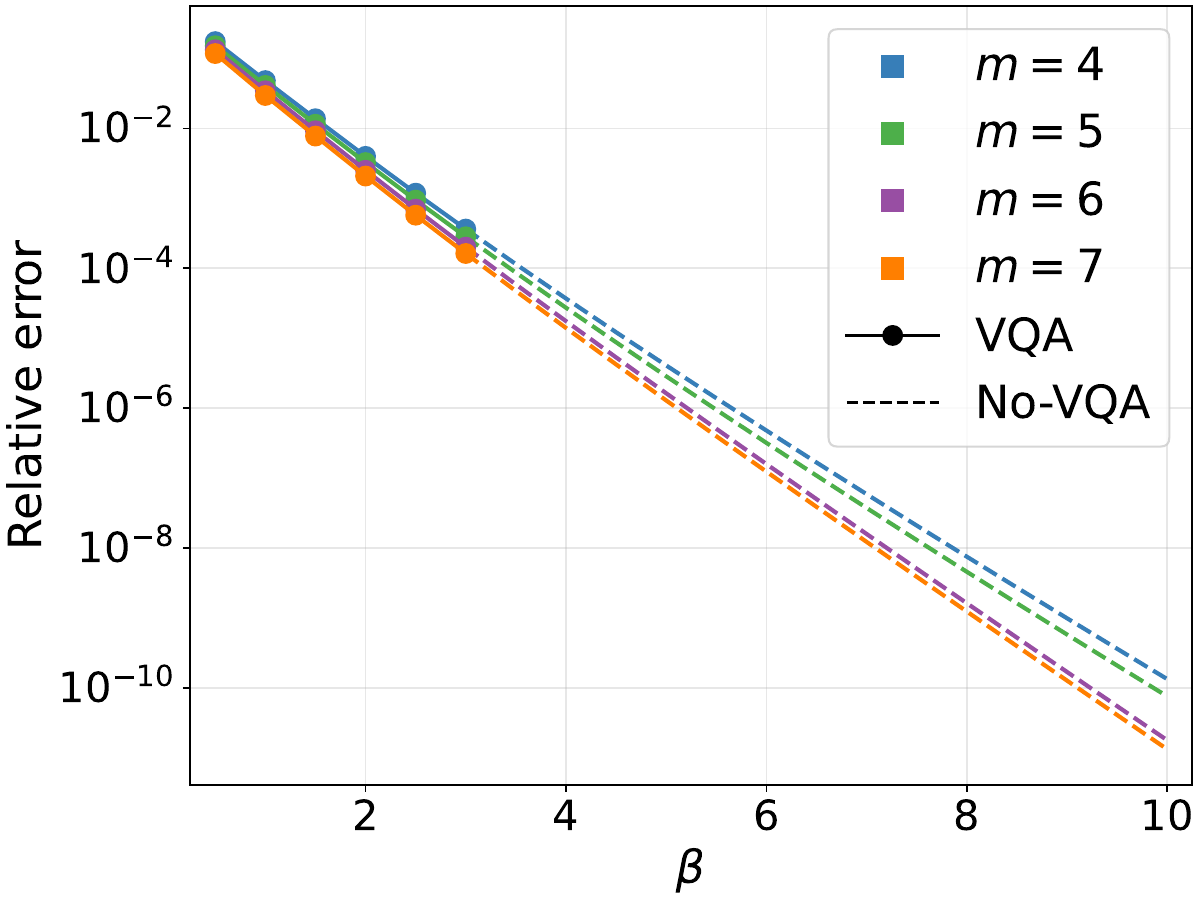}
    \caption{The relative free energy error as a function of $\beta$ for the XXZ model with $N_s=6$ and $J_z=-1.5$. For $\beta\leq\beta_0$, we prepare the Gibbs state using TEPID-ADAPT, as indicated by the markers. For $\beta>\beta_0$, the same converged adaptive unitary is used with no parameter re-optimization, as indicated by the dashed lines.}
    \label{fig:FErr_Tmscan_Ferro}
\end{figure}

For each of these phases, we show two figures that compare our results with exact diagonalization:
\begin{itemize}
    \item Figs.~\ref{fig:Infids_mscan_Ferro},~\ref{fig:Infids_mscan_Para},~\ref{fig:Infids_mscan_AFerro} compare the states prepared using TEPID-ADAPT with exact diagonalization for a fixed temperature. We show the infidelity (left axis) and relative errors (right axis) of the energies of the eigenstates in the truncated space, and the free energy of both the truncated $m-$rank Gibbs state and the full Gibbs state. The numbered state indices on the horizontal axes of these plots correspond to the eigenstates of the Hamiltonian. These eigenstates are prepared using Eq.~\eqref{eq:EigPrep}, and their energies are measured using Eq.~\eqref{eq:EnergyAccess}. The last index $\rho_G$ corresponds to the Gibbs state, and its relative error refers to the free energy in Eq.~\eqref{eq:FEnergy}. We show results for various values of the cutoff $m$, as indicated by the colors in the legends.
    
    \item In Figs.~\ref{fig:FErr_Tmscan_Ferro},~\ref{fig:FErr_Tmscan_Para},~\ref{fig:FErr_Tmscan_AFerro}, we demonstrate the ability of TEPID-ADAPT to prepare Gibbs states at lower temperatures ($\beta>\beta_0$). We plot the relative free-energy errors of the prepared Gibbs state with the exact one. We run the VQA for $\beta\leq\beta_0$, as indicated by the markers. For $\beta>\beta_0$, we use the same \textit{converged} adaptive unitary $V_A(\vec{\theta}^*)$ on the system qubits with a modified set of parameters in $U_m$, following Eq.~\eqref{eq:ModifiedStatePrepParams}.
\end{itemize}
The relative error of a quantity $Q$ is defined as
\begin{equation}
    \varepsilon_{\text{rel}}(Q) = \left\vert\frac{Q-Q_{\text{ex}}}{Q_{\text{ex}}}\right\vert,
\end{equation}
where the subscript ex corresponds to results from exact diagonalization.

The definition of fidelity between two density matrices $\rho,\sigma$ we use in this article is
\begin{equation}
    \mathcal{F}(\rho,\sigma)=\left(\Tr\sqrt{\sqrt{\rho}\sigma\sqrt{\rho}}\right)^2,
\end{equation}
which reduces to 
\begin{equation}
    \mathcal{F}(\ket{\psi},\ket{\phi})=\lvert\bra{\phi}\ket{\psi}\rvert^2
\end{equation}
for pure states $\ket{\psi},\ket{\phi}$. In the case of degeneracies, we find the fidelity of the prepared state with the degenerate subspace
\begin{equation}
    \mathcal{F}\left({\ket{\psi},\{\ket{\phi_j}\}_{j=1}^D}\right)=\sum_{j=1}^D\lvert\bra{\psi}\ket{\phi_j}\rvert^2,
\end{equation}
where $\{\ket{\phi_j}\}_{j=1}^D$ is a set of orthonormal vectors that spans the $D$-degenerate subspace.

In Appendix~\ref{App:Tolerances}, we explore the effect of gradient tolerances on the convergence of TEPID-ADAPT. We find qualitative differences in the convergence of the free energy for the different phases of the XXZ model. We analyze these differences and relate them to the structure of the low energy spectrum of the model. We also include a numerical study of how the cutoff $m$ scales with $N_s$ for a given temperature and fidelity threshold for the Heisenberg XXZ model in Appendix~\ref{App:Scaling}. For translationally invariant, local spin systems, we expect the rank $m$ to asymptotically scale exponentially with $N_s$ for a fixed fidelity threshold. We numerically find that this expectation holds true for tight fidelity thresholds, even at moderate system sizes, $N_s$. However, upon relaxing the fidelity thresholds, we numerically find that the scaling could follow slower growths at moderate system sizes.

%%%%%%%%%%%%%%%%%%%%%%%%%%%%%%%%%%%%%%%%%%%%%%%%%%%%%%%%%%%%%%%%%%%%%%%%%%%%%%%%%%%%%%%%%%

\subsection{Ferromagnetic phase $(J_z<-1)$}\label{subsec:Ferro}

\begin{figure}[ht!]
    \centering
    \includegraphics[width=\linewidth]{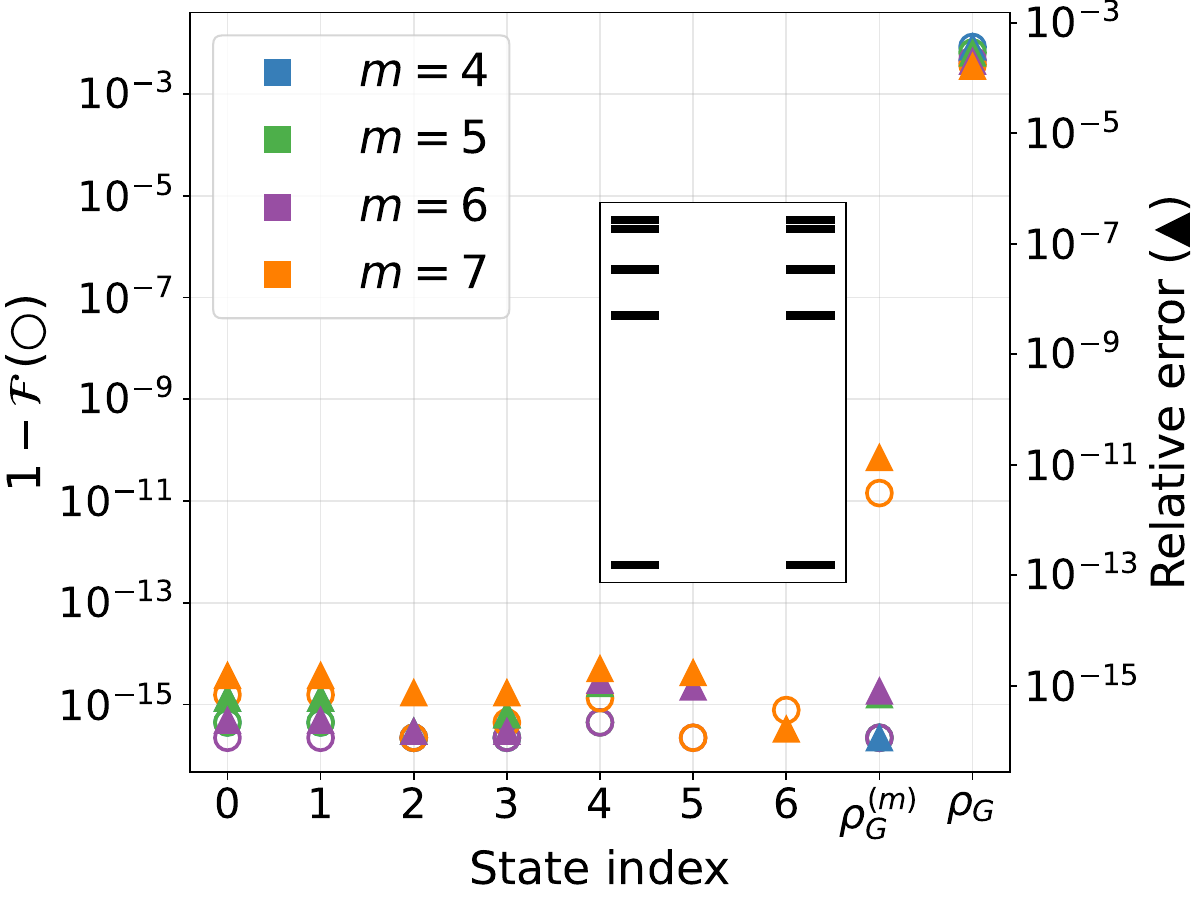}
    \caption{The infidelities (left axis, circle markers) and relative errors (right axis, triangle markers) of the low-lying eigenenergies, and of the free energy of the $\beta_0=3.0$ Gibbs state (truncated and full) of the XXZ model with $N_s=6$ and $J_z=-1.5$. The numbered state indices refer to the eigenstates, $\rho_G$ to the Gibbs state, and $\rho^{(m)}_G$ is the truncated $m$-rank Gibbs state. The inset shows the low-lying eigenspectrum.}
    \label{fig:Infids_mscan_Ferro}
\end{figure}

In the ferromagnetic phase, it is energetically favorable for the spins of the Heisenberg chain to be aligned. As a result, we consider the following computational subspace spanned by states with spins that are fully aligned, or with a low degree of misalignment near the boundaries, as we expect these excitations to incur the smallest energy penalties:

\begin{multline}
    \left\{c_k\right\}\equiv\left\{\ket{000000},\ket{111111},\ket{000001},\ket{111110},\right.\\
 \left.\ket{011111}, \ket{100000}, \ket{001111}\right\}.   
\end{multline}
In this phase, the ground state is degenerate with a sizable gap. The low-lying eigenspectrum above the degenerate ground state subspace is dense, as shown in the inset of Fig.~\ref{fig:Infids_mscan_Ferro}. As a consequence, including a small amount of these states does not significantly improve the fidelity of the Gibbs state, as seen in Fig.~\ref{fig:Infids_mscan_Ferro}.

%%%%%%%%%%%%%%%%%%%%%%%%%%%%%%%%%%%%%%%%%%%%%%%%%%%%%%%%%%%%%%%%%%%%%%%%%%%%%%%%%%%%%%%%%%

\subsection{Paramagnetic phase ($-1<J_z<1$)}\label{subsec:Para}

\begin{figure}[ht!]
    \centering
    \includegraphics[width=\linewidth]{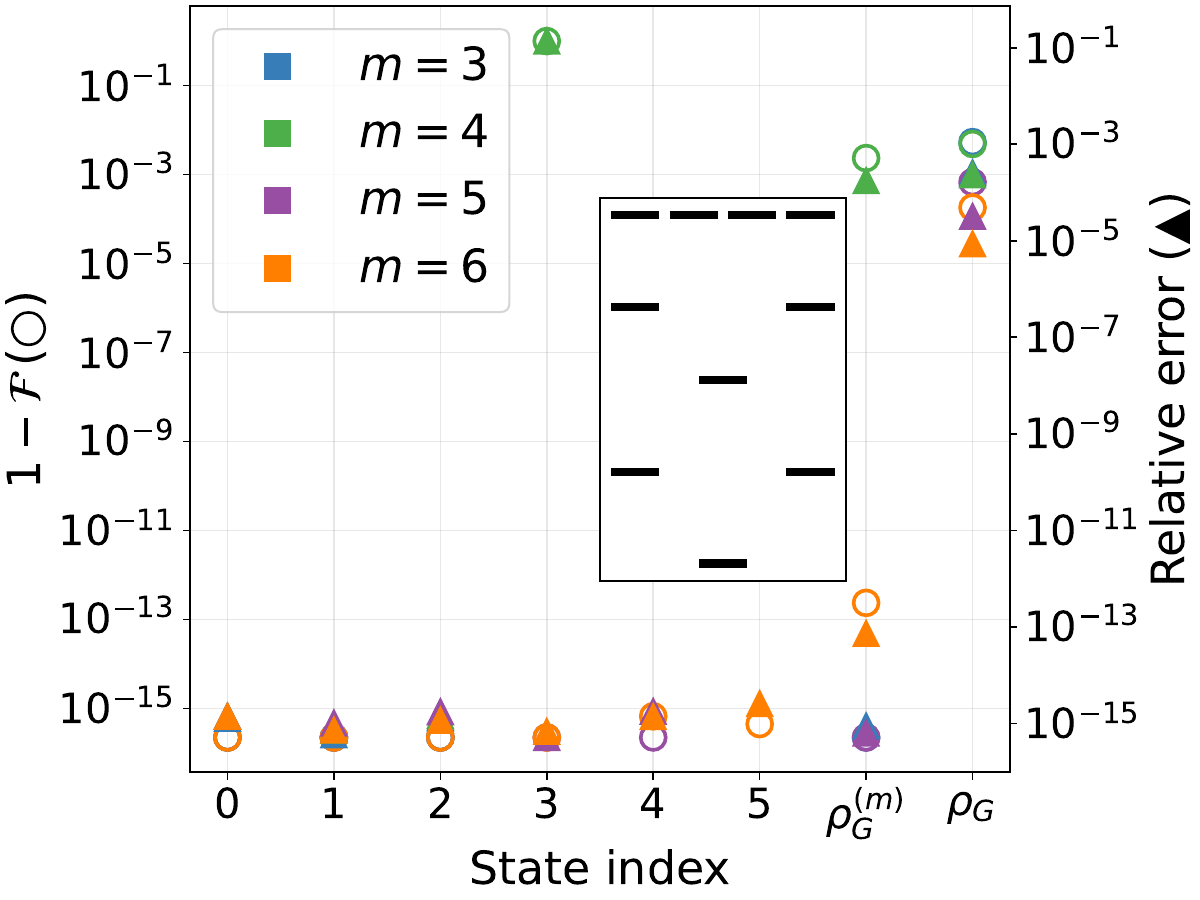}
    \caption{The infidelities (left axis, circle markers) and relative errors (right axis, triangle markers) of the low-lying eigenenergies, and of the free energy of the $\beta_0=3.0$ Gibbs state (truncated and full) of the XXZ model with $N_s=6$ and $J_z=0.0$. The numbered state indices refer to the eigenstates, $\rho_G$ to the Gibbs state, and $\rho^{(m)}_G$ is the truncated $m$-rank Gibbs state. The inset shows the low-lying eigenspectrum.}
    \label{fig:Infids_mscan_Para}
\end{figure}

\begin{figure}[ht!]
    \centering
    \includegraphics[width=\linewidth]{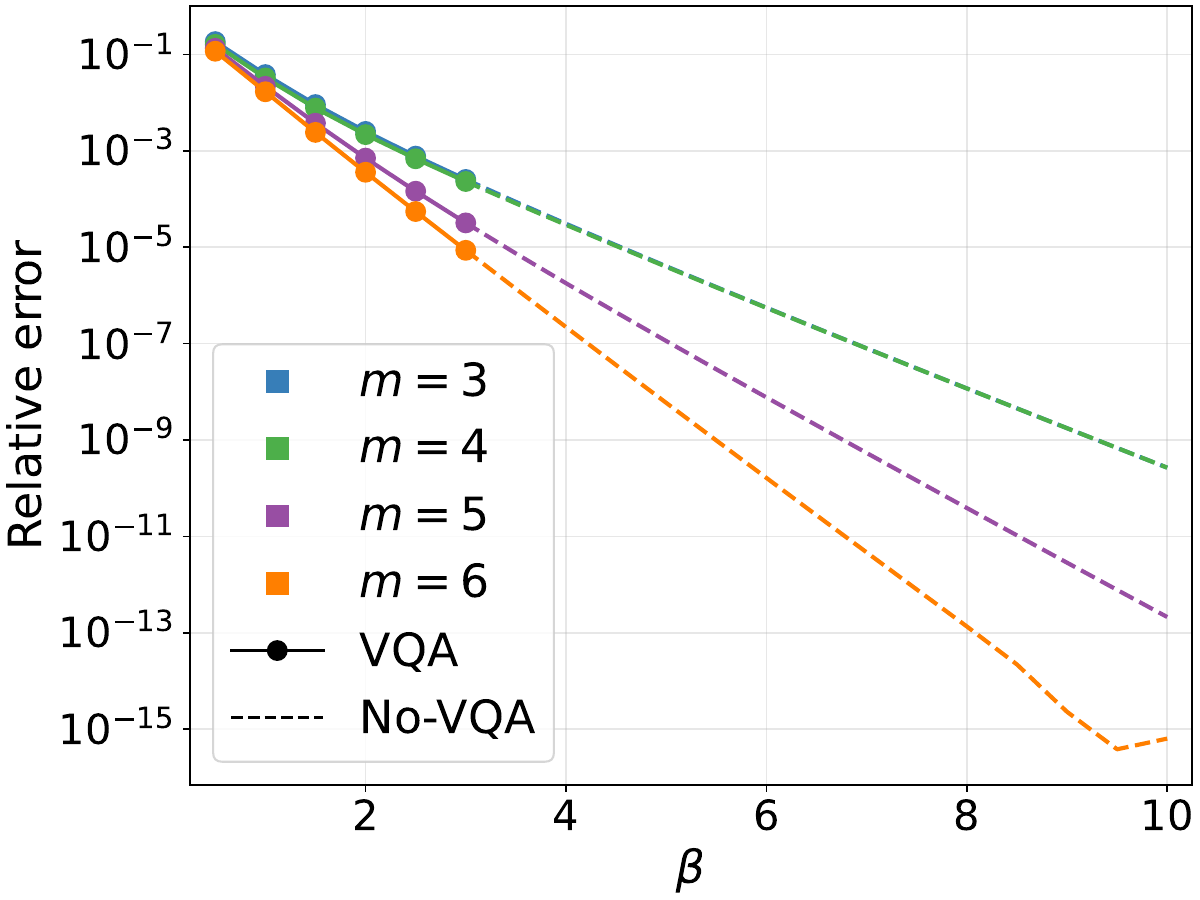}
    \caption{The relative free energy error as a function of $\beta$ for the XXZ model with $N_s=6$ and $J_z=0.0$. For $\beta\leq\beta_0$, we prepare the Gibbs state using TEPID-ADAPT, as indicated by the markers. For $\beta>\beta_0$, the same converged adaptive unitary is used with no parameter re-optimization, as indicated by the dashed lines.}
    \label{fig:FErr_Tmscan_Para}
\end{figure}

We have chosen $J_z=0$, where the model reduces to the XY spin chain. In the paramagnetic phase, in the absence of an external magnetic field, the spins average to zero net magnetization. As a result, we consider the following computational subspace that includes states with zero or low net magnetization ($M\sim\sum_j\langle\sigma^z_j\rangle$)

\begin{multline}
    \left\{c_k\right\}\equiv\left\{\ket{010101},\ket{010111},\ket{010100},\ket{100001},\right.\\
 \left. \ket{010110},\ket{011110}\right\}.    
\end{multline}
For $m=4$, we find that the third excited state is not correctly prepared, as seen in Figs.~\ref{fig:Infids_mscan_Para},~\ref{fig:FErr_Tmscan_Para}. (We find that it instead prepares one of the higher excited states.) However, for larger values of $m$, this is not the case. This emphasizes the need to prepare the Gibbs state using different values of the cutoff $m$ to benchmark the reliability of TEPID-ADAPT in finding the lowest eigenstates. We investigate this in further detail in Appendix~\ref{App:Paramagnetic3state}.

%%%%%%%%%%%%%%%%%%%%%%%%%%%%%%%%%%%%%%%%%%%%%%%%%%%%%%%%%%%%%%%%%%%%%%%%%%%%%%%%%%%%%%%%%%

\subsection{Antiferromagnetic phase ($J_z>1$)}\label{subsec:AntiFerro}

\begin{figure}[ht!]
    \centering
    \includegraphics[width=\linewidth]{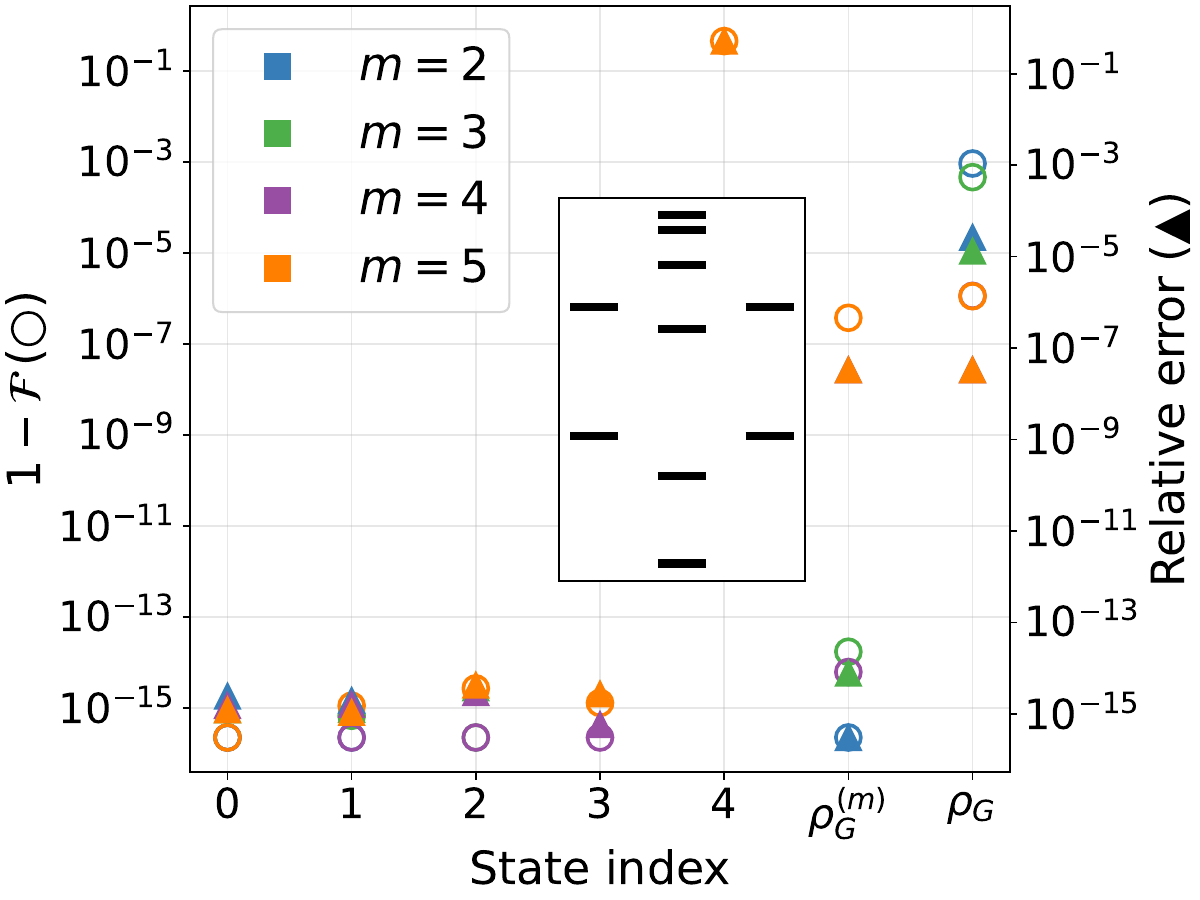}
    \caption{The infidelities (left axis, circle markers) and relative errors (right axis, triangle markers) of the low-lying eigenenergies, and of the free energy of the $\beta_0=3.0$ Gibbs state (truncated and full) of the XXZ model with $N_s=6$ and $J_z=1.5$. The numbered state indices refer to the eigenstates, $\rho_G$ to the Gibbs state, and $\rho^{(m)}_G$ is the truncated $m$-rank Gibbs state. The inset shows the low-lying eigenspectrum.}
    \label{fig:Infids_mscan_AFerro}
\end{figure}

\begin{figure}[ht!]
    \centering
    \includegraphics[width=\linewidth]{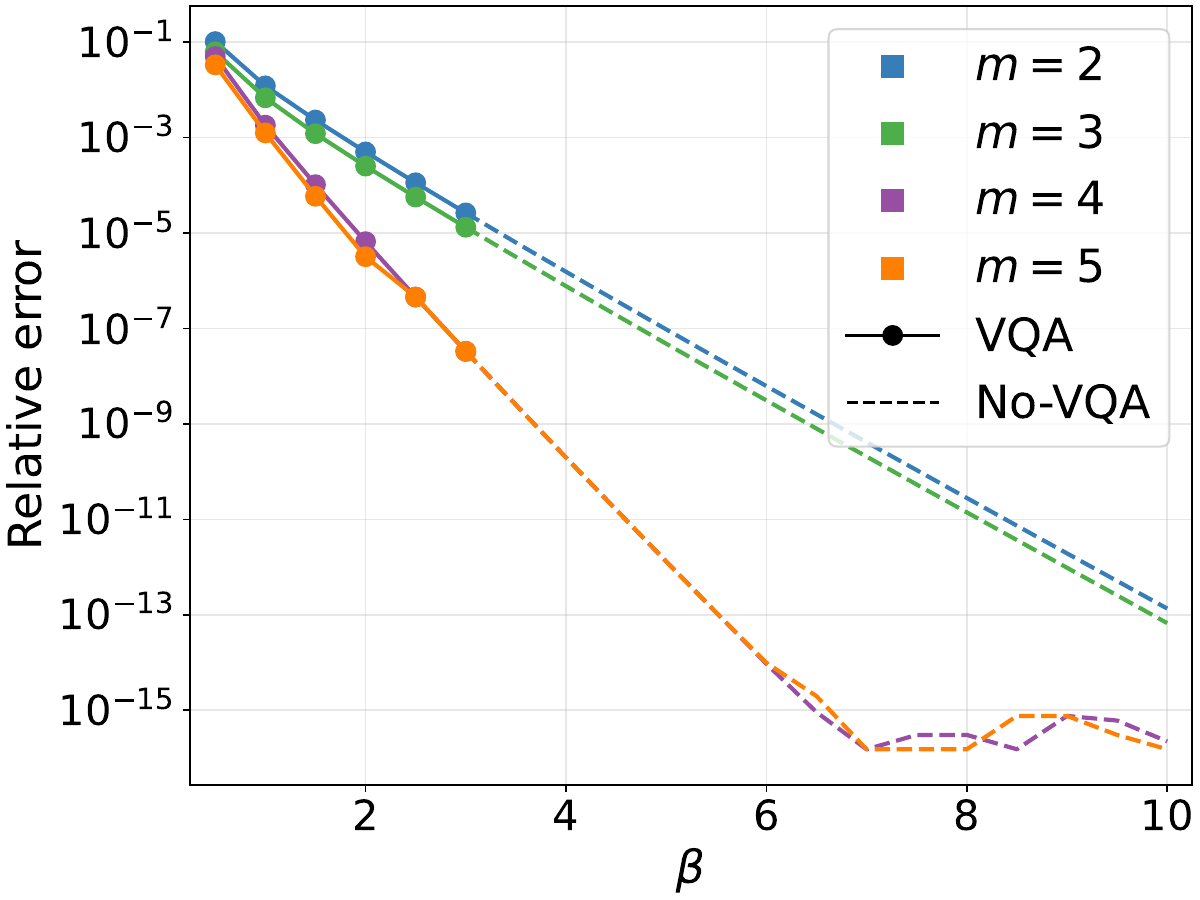}
    \caption{The relative free energy error as a function of $\beta$ for the XXZ model with $N_s=6$ and $J_z=1.5$. For $\beta\leq\beta_0$, we prepare the Gibbs state using TEPID-ADAPT, as indicated by the markers. For $\beta>\beta_0$, the same converged adaptive unitary is used with no parameter re-optimization, as indicated by the dashed lines.}
    \label{fig:FErr_Tmscan_AFerro}
\end{figure}

In the antiferromagnetic phase, it is energetically favorable for the spins of the Heisenberg chain to be antialigned. As a result, we consider the following computational subspace spanned by states that are either fully, or mostly antialigned 
\begin{multline}
    \left\{c_k\right\}\equiv\left\{\ket{010101},\ket{010110},\ket{010100},\ket{010111},\right.\\
 \left. \ket{100101}\right\}.
\end{multline}
In this phase, the low-lying eigenspectrum is sparse, as shown in the inset of Fig.~\ref{fig:Infids_mscan_AFerro}. This allows us to achieve high fidelities with a relatively small $m$. For $m=5$ in Fig.~\ref{fig:Infids_mscan_AFerro}, despite the failure to prepare the fourth excited state, the fidelity of the Gibbs state is high. This shows that this state does not contribute significantly to the Gibbs state at this temperature. This is also evidenced by Fig.~\ref{fig:FErr_Tmscan_AFerro}, where for $\beta\leq2.0$, the fourth excited state is successfully prepared, indicated by the marginally lower relative free energy error. 

%%%%%%%%%%%%%%%%%%%%%%%%%%%%%%%%%%%%%%%%%%%%%%%%%%%%%%%%%%%%%%%%%%%%%%%%%%%%%%%%%%%%%%%%%%
%%%%%%%%%%%%%%%%%%%%%%%%%%%%%%%%%%%%%%%%%%%%%%%%%%%%%%%%%%%%%%%%%%%%%%%%%%%%%%%%%%%%%%%%%%

\section{Conclusions and outlook}\label{sec:Conclusions}

In this article, we introduced TEPID-ADAPT, a variational quantum method to simultaneously prepare low-temperature Gibbs states and the corresponding low-energy eigenstates. We constructed a modular ansatz that is partially static and partially adaptive, and uses a minimal number of ancillary qubits for state purification. The static part spans the extended qubit register, and prepares a parametrized density matrix on the system register. This density matrix is diagonal in the computational basis. The adaptive part of the ansatz has support on only the system register. It aims to find a unitary that rotates the computational basis subspace to the truncated eigenspace, thereby approximately rotating the prepared density matrix to the Gibbs state. This approximation is good at low temperatures, where only the low-lying states contribute significantly to the Gibbs state, provided that the initially chosen computational subspace is large enough. 

The nature of the adaptive part of the ansatz grants another nice feature. At temperatures lower than that of the prepared Gibbs state, the excited states become increasingly less important. As a result, we are able to use the same converged adaptive ansatz to prepare lower temperature Gibbs states \textit{without} any parameter re-optimization. We do have to change the parameters in the static portion of the ansatz appropriately.

The access to the low-lying eigenstates also enables the preparation of low-temperature TFD states without any parameter optimization. While by definition we do need to double the number of system qubits, we do not perform variational optimization on this expanded system. We simply use the converged adaptively generated basis-change unitary on both halves of the expanded system to prepare the TFD state.

A crucial part of VQE is the choice of reference state. This typically has a bearing on the depth of the circuit found to get to the target state. This holds true for TEPID-ADAPT as well --- it is important to choose a set of computational basis elements that are effective at finding the low-lying eigenstates that make up the target Gibbs state. In this work, we use system symmetries, and the known phase structure of the Heisenberg XXZ model to make this choice. However, for other models where the choice is less clear, it is important to have a systematic procedure to choose the computational subspace. We save this exploration for future work.

The choice of an operator pool in ADAPT also has a bearing on its effectiveness. In this work, we used the set of all Pauli strings on the system qubits as the pool. The size of this pool scales exponentially in the size of the system. For example, a better choice of an operator pool would take advantage of system symmetries. This has been found to be effective for ground state preparation using ADAPT-VQE~\cite{VanDykePRR2024, FarrellPRXQ2024}. We save the analysis of choice of operator pools for TEPID-ADAPT for future work.

In this work, the results were obtained using noise-free state vector simulations. Analyzing the effects of shot noise in estimating the cost function and gradients is important. Similarly, studying the effects of device noise using commonly used noise models/channels is also important, especially in light of the fact that we are variationally optimizing over mixed states. This would also be a good setting to compare the two implementations of our algorithm---with and without ancillary qubits. We save these for future work.

A generalization of our method that one might consider is to allow $m$ to be large and let $\mu$ be a parameterized family of distributions over $[m]$, such as a product distribution or a classical Gibbs distribution. We might choose a smaller set of parameters $\vec{\nu} = (\nu_1,\ldots,\nu_\ell)$ and let $\vec\mu \equiv \vec\mu(\vec \nu)$. In this case, we would need an efficient gate decomposition of $U_m(\vec{\mu})$, or equivalently an efficient way to sample computational basis states according to the distribution $\vec{\mu}(\vec{\nu})$. We also would need to retain the ability to analytically compute the entropy and the cost-function gradients using $\vec{\nu}$ or the gate parameters in the decomposition of $U_m(\vec{\mu}(\vec{\nu}))$.

In this work, we show how the cutoff $m$ scales with the system size for the Heisenberg XXZ model in Appendix~\ref{App:Scaling}. A more general treatment that is less dependent on the particular model at hand is warranted. We save this for future work as well. 

%%%%%%%%%%%%%%%%%%%%%%%%%%%%%%%%%%%%%%%%%%%%%%%%%%%%%%%%%%%%%%%%%%%%%%%%%%%%%%%%%%%%%%%%%%
%%%%%%%%%%%%%%%%%%%%%%%%%%%%%%%%%%%%%%%%%%%%%%%%%%%%%%%%%%%%%%%%%%%%%%%%%%%%%%%%%%%%%%%%%%

\section*{Acknowledgments}\label{sec:Thanks}

Bharath thanks Arshag Danageozian for helpful discussions on quantum channels, Jason Pollack for helpful discussions on unitary equivalence, Christopher K Long and Abhishek Kumar for useful feedback during the development of TEPID-ADAPT. S. E. Economou thanks Pablo Sala for an interesting question, which led to us proving and including Theorem 1 in the second version of the paper.
The Quantikz package ~\cite{Kay2023} was used to generate the circuits in this paper.

B. Sambasivam, K. Sherbert, K. Shirali, A. Harrow, and S. E. Economou are supported by the U.S. Department of Energy, Office of Science, National Quantum Information Science Research Centers, Co-design Center for Quantum Advantage (C$^2$QA) under Contract No.DE-SC0012704. N. J. Mayhall acknowledges funding from the Department of Energy (Award number: DE-SC0024619). E. Barnes acknowledges support from the Department of Energy (Award no. DE-SC0025430).

%%%%%%%%%%%%%%%%%%%%%%%%%%%%%%%%%%%%%%%%%%%%%%%%%%%%%%%%%%%%%%%%%%%%%%%%%%%%%%%%%%%%%%%%%%
%%%%%%%%%%%%%%%%%%%%%%%%%%%%%%%%%%%%%%%%%%%%%%%%%%%%%%%%%%%%%%%%%%%%%%%%%%%%%%%%%%%%%%%%%%

\section*{Author contributions}\label{sec:Contributions}

B. Sambasivam conceived the project, developed the main methodology, implemented most of the computational code, performed the calculations, and wrote the initial draft of the manuscript. K. Sherbert and K. Shirali assisted with code development and calculations. N. J. Mayhall, A. W. Harrow, E. Barnes, and S. E. Economou contributed to improving and extending the project through conceptual input and analysis of the main results. E. Barnes proved Theorem 1. All authors discussed the results, contributed to refining the manuscript, and approved the final version.

%%%%%%%%%%%%%%%%%%%%%%%%%%%%%%%%%%%%%%%%%%%%%%%%%%%%%%%%%%%%%%%%%%%%%%%%%%%%%%%%%%%%%%%%%%
%%%%%%%%%%%%%%%%%%%%%%%%%%%%%%%%%%%%%%%%%%%%%%%%%%%%%%%%%%%%%%%%%%%%%%%%%%%%%%%%%%%%%%%%%%

\bibliography{apssamp}% Produces the bibliography via BibTeX.

@article{ChildsPNAS2018,
author = {Andrew M. Childs  and Dmitri Maslov  and Yunseong Nam  and Neil J. Ross  and Yuan Su },
title = {Toward the first quantum simulation with quantum speedup},
journal = {Proceedings of the National Academy of Sciences},
volume = {115},
number = {38},
pages = {9456-9461},
year = {2018},
doi = {10.1073/pnas.1801723115},
URL = {https://www.pnas.org/doi/abs/10.1073/pnas.1801723115},
eprint = {https://www.pnas.org/doi/pdf/10.1073/pnas.1801723115}}

@article{Garcia-Mata2023,
   title={Out-of-time-order correlations and quantum chaos},
   volume={18},
   ISSN={1941-6016},
   url={http://dx.doi.org/10.4249/scholarpedia.55237},
   DOI={10.4249/scholarpedia.55237},
   number={4},
   journal={Scholarpedia},
   publisher={Scholarpedia},
   author={García-Mata, Ignacio and Jalabert, Rodolfo and Wisniacki, Diego},
   year={2023},
   pages={55237} }

@article{KirkpatrickScience1983,
author = {S. Kirkpatrick  and C. D. Gelatt  and M. P. Vecchi },
title = {Optimization by Simulated Annealing},
journal = {Science},
volume = {220},
number = {4598},
pages = {671-680},
year = {1983},
doi = {10.1126/science.220.4598.671},
URL = {https://www.science.org/doi/abs/10.1126/science.220.4598.671},}

@article{SommaPRL2008,
  title = {Quantum Simulations of Classical Annealing Processes},
  author = {Somma, R. D. and Boixo, S. and Barnum, H. and Knill, E.},
  journal = {Phys. Rev. Lett.},
  volume = {101},
  issue = {13},
  pages = {130504},
  numpages = {4},
  year = {2008},
  month = {Sep},
  publisher = {American Physical Society},
  doi = {10.1103/PhysRevLett.101.130504},
  url = {https://link.aps.org/doi/10.1103/PhysRevLett.101.130504}
}

@article{KieferovaPRA2017,
  title = {Tomography and generative training with quantum Boltzmann machines},
  author = {Kieferov\'a, M\'aria and Wiebe, Nathan},
  journal = {Phys. Rev. A},
  volume = {96},
  issue = {6},
  pages = {062327},
  numpages = {13},
  year = {2017},
  month = {Dec},
  publisher = {American Physical Society},
  doi = {10.1103/PhysRevA.96.062327},
  url = {https://link.aps.org/doi/10.1103/PhysRevA.96.062327}
}

@article{BiamonteNature2017,
  title = {Quantum machine learning},
  volume = {549},
  ISSN = {1476-4687},
  url = {http://dx.doi.org/10.1038/nature23474},
  DOI = {10.1038/nature23474},
  number = {7671},
  journal = {Nature},
  publisher = {Springer Science and Business Media LLC},
  author = {Biamonte,  Jacob and Wittek,  Peter and Pancotti,  Nicola and Rebentrost,  Patrick and Wiebe,  Nathan and Lloyd,  Seth},
  year = {2017},
  month = sep,
  pages = {195–202}
}

@misc{Watrous2008,
      title={Quantum Computational Complexity}, 
      author={John Watrous},
      year={2008},
      eprint={0804.3401},
      archivePrefix={arXiv},
      primaryClass={quant-ph},
      url={https://arxiv.org/abs/0804.3401}, 
}

@article{AharanovACM2013,
author = {Aharonov, Dorit and Arad, Itai and Vidick, Thomas},
title = {Guest column: the quantum PCP conjecture},
year = {2013},
issue_date = {June 2013},
publisher = {Association for Computing Machinery},
address = {New York, NY, USA},
volume = {44},
number = {2},
issn = {0163-5700},
url = {https://doi.org/10.1145/2491533.2491549},
doi = {10.1145/2491533.2491549},
journal = {SIGACT News},
month = jun,
pages = {47–79},
numpages = {33}
}

@INPROCEEDINGS{BakshiIEEE2024,
  author={Bakshi, Ainesh and Liu, Allen and Moitra, Ankur and Tang, Ewin},
  booktitle={2024 IEEE 65th Annual Symposium on Foundations of Computer Science (FOCS)}, 
  title={High-Temperature Gibbs States are Unentangled and Efficiently Preparable}, 
  year={2024},
  volume={},
  number={},
  pages={1027-1036},
  doi={10.1109/FOCS61266.2024.00068}}

@article{PoulinPRL2009,
  title = {Sampling from the Thermal Quantum Gibbs State and Evaluating Partition Functions with a Quantum Computer},
  author = {Poulin, David and Wocjan, Pawel},
  journal = {Phys. Rev. Lett.},
  volume = {103},
  issue = {22},
  pages = {220502},
  numpages = {4},
  year = {2009},
  month = {Nov},
  publisher = {American Physical Society},
  doi = {10.1103/PhysRevLett.103.220502},
  url = {https://link.aps.org/doi/10.1103/PhysRevLett.103.220502}
}

@article{TemmeNature2011,
  title = {Quantum Metropolis sampling},
  volume = {471},
  ISSN = {1476-4687},
  url = {http://dx.doi.org/10.1038/nature09770},
  DOI = {10.1038/nature09770},
  number = {7336},
  journal = {Nature},
  publisher = {Springer Science and Business Media LLC},
  author = {Temme,  K. and Osborne,  T. J. and Vollbrecht,  K. G. and Poulin,  D. and Verstraete,  F.},
  year = {2011},
  month = mar,
  pages = {87–90}
}

@article{ChowdhuryQIC2017,
author = {Chowdhury, Anirban Narayan and Somma, Rolando D.},
title = {Quantum algorithms for Gibbs sampling and hitting-time estimation},
year = {2017},
issue_date = {February 2017},
publisher = {Rinton Press, Incorporated},
address = {Paramus, NJ},
volume = {17},
number = {1–2},
issn = {1533-7146},
journal = {Quantum Info. Comput.},
month = feb,
pages = {41–64},
numpages = {24}
}

@ARTICLE{EassaNPJQI2024,
  title     = "Gibbs state sampling via cluster expansions",
  author    = "Eassa, Norhan M and Moustafa, Mahmoud M and Banerjee, Arnab and
               Cohn, Jeffrey",
  journal   = "Npj Quantum Inf.",
  publisher = "Springer Science and Business Media LLC",
  volume    =  10,
  number    =  1,
  month     =  oct,
  year      =  2024,
  copyright = "https://creativecommons.org/licenses/by-nc-nd/4.0",
  language  = "en"
}

@article{MottaNaturePhys2019,
  title = {Determining eigenstates and thermal states on a quantum computer using quantum imaginary time evolution},
  volume = {16},
  ISSN = {1745-2481},
  url = {http://dx.doi.org/10.1038/s41567-019-0704-4},
  DOI = {10.1038/s41567-019-0704-4},
  number = {2},
  journal = {Nature Physics},
  publisher = {Springer Science and Business Media LLC},
  author = {Motta,  Mario and Sun,  Chong and Tan,  Adrian T. K. and O’Rourke,  Matthew J. and Ye,  Erika and Minnich,  Austin J. and Brandão,  Fernando G. S. L. and Chan,  Garnet Kin-Lic},
  year = {2019},
  month = nov,
  pages = {205–210}
}

@article{SunPRXQ2021,
  title = {Quantum Computation of Finite-Temperature Static and Dynamical Properties of Spin Systems Using Quantum Imaginary Time Evolution},
  author = {Sun, Shi-Ning and Motta, Mario and Tazhigulov, Ruslan N. and Tan, Adrian T.K. and Chan, Garnet Kin-Lic and Minnich, Austin J.},
  journal = {PRX Quantum},
  volume = {2},
  issue = {1},
  pages = {010317},
  numpages = {14},
  year = {2021},
  month = {Feb},
  publisher = {American Physical Society},
  doi = {10.1103/PRXQuantum.2.010317},
  url = {https://link.aps.org/doi/10.1103/PRXQuantum.2.010317}
}

@article{KamakariPRXQ2022,
  title = {Digital Quantum Simulation of Open Quantum Systems Using Quantum Imaginary--Time Evolution},
  author = {Kamakari, Hirsh and Sun, Shi-Ning and Motta, Mario and Minnich, Austin J.},
  journal = {PRX Quantum},
  volume = {3},
  issue = {1},
  pages = {010320},
  numpages = {10},
  year = {2022},
  month = {Feb},
  publisher = {American Physical Society},
  doi = {10.1103/PRXQuantum.3.010320},
  url = {https://link.aps.org/doi/10.1103/PRXQuantum.3.010320}
}

@Article{GetelinaSciPostPhys2023,
	title={{Adaptive variational quantum minimally entangled typical thermal states for finite temperature simulations}},
	author={João C. Getelina and Niladri Gomes and Thomas Iadecola and Peter P. Orth and Yong-Xin Yao},
	journal={SciPost Phys.},
	volume={15},
	pages={102},
	year={2023},
	publisher={SciPost},
	doi={10.21468/SciPostPhys.15.3.102},
	url={https://scipost.org/10.21468/SciPostPhys.15.3.102},
}

@article{MaldacenaJHEP2003,
doi = {10.1088/1126-6708/2003/04/021},
url = {https://dx.doi.org/10.1088/1126-6708/2003/04/021},
year = {2003},
month = {apr},
publisher = {},
volume = {2003},
number = {04},
pages = {021},
author = {Juan Maldacena},
title = {Eternal black holes in anti-de Sitter},
journal = {Journal of High Energy Physics}
}

@article{MaldacenaFdP2013,
   title={Cool horizons for entangled black holes},
   volume={61},
   ISSN={1521-3978},
   url={http://dx.doi.org/10.1002/prop.201300020},
   DOI={10.1002/prop.201300020},
   number={9},
   journal={Fortschritte der Physik},
   publisher={Wiley},
   author={Maldacena, J. and Susskind, L.},
   year={2013},
   month=aug, pages={781–811} 
}

@article{CottrellJHEP2019,
  title = {How to build the thermofield double state},
  volume = {2019},
  ISSN = {1029-8479},
  url = {http://dx.doi.org/10.1007/JHEP02(2019)058},
  DOI = {10.1007/jhep02(2019)058},
  number = {2},
  journal = {Journal of High Energy Physics},
  publisher = {Springer Science and Business Media LLC},
  author = {Cottrell,  William and Freivogel,  Ben and Hofman,  Diego M. and Lokhande,  Sagar F.},
  year = {2019},
  month = feb 
}

@article{WuPRL2019,
  title = {Variational Thermal Quantum Simulation via Thermofield Double States},
  author = {Wu, Jingxiang and Hsieh, Timothy H.},
  journal = {Phys. Rev. Lett.},
  volume = {123},
  issue = {22},
  pages = {220502},
  numpages = {6},
  year = {2019},
  month = {Nov},
  publisher = {American Physical Society},
  doi = {10.1103/PhysRevLett.123.220502},
  url = {https://link.aps.org/doi/10.1103/PhysRevLett.123.220502}
}

@article{SagastizabalNPJQI2021,
  title = {Variational preparation of finite-temperature states on a quantum computer},
  volume = {7},
  ISSN = {2056-6387},
  url = {http://dx.doi.org/10.1038/s41534-021-00468-1},
  DOI = {10.1038/s41534-021-00468-1},
  number = {1},
  journal = {npj Quantum Information},
  publisher = {Springer Science and Business Media LLC},
  author = {Sagastizabal,  R. and Premaratne,  S. P. and Klaver,  B. A. and Rol,  M. A. and Negîrneac,  V. and Moreira,  M. S. and Zou,  X. and Johri,  S. and Muthusubramanian,  N. and Beekman,  M. and Zachariadis,  C. and Ostroukh,  V. P. and Haider,  N. and Bruno,  A. and Matsuura,  A. Y. and DiCarlo,  L.},
  year = {2021},
  month = aug 
}

@article{WangPRApp2021,
  title = {Variational Quantum Gibbs State Preparation with a Truncated Taylor Series},
  author = {Wang, Youle and Li, Guangxi and Wang, Xin},
  journal = {Phys. Rev. Appl.},
  volume = {16},
  issue = {5},
  pages = {054035},
  numpages = {17},
  year = {2021},
  month = {Nov},
  publisher = {American Physical Society},
  doi = {10.1103/PhysRevApplied.16.054035},
  url = {https://link.aps.org/doi/10.1103/PhysRevApplied.16.054035}
}

@misc{Warren2022,
      title={Adaptive variational algorithms for quantum Gibbs state preparation}, 
      author={Ada Warren and Linghua Zhu and Nicholas J. Mayhall and Edwin Barnes and Sophia E. Economou},
      year={2022},
      eprint={2203.12757},
      archivePrefix={arXiv},
      primaryClass={quant-ph},
      url={https://arxiv.org/abs/2203.12757}, 
}

@article{ConsiglioPRA2024,
  title = {Variational Gibbs state preparation on noisy intermediate-scale quantum devices},
  author = {Consiglio, Mirko and Settino, Jacopo and Giordano, Andrea and Mastroianni, Carlo and Plastina, Francesco and Lorenzo, Salvatore and Maniscalco, Sabrina and Goold, John and Apollaro, Tony J. G.},
  journal = {Phys. Rev. A},
  volume = {110},
  issue = {1},
  pages = {012445},
  numpages = {14},
  year = {2024},
  month = {Jul},
  publisher = {American Physical Society},
  doi = {10.1103/PhysRevA.110.012445},
  url = {https://link.aps.org/doi/10.1103/PhysRevA.110.012445}
}

@article{KandalaNature2017,
  title = {Hardware-efficient variational quantum eigensolver for small molecules and quantum magnets},
  volume = {549},
  ISSN = {1476-4687},
  url = {http://dx.doi.org/10.1038/nature23879},
  DOI = {10.1038/nature23879},
  number = {7671},
  journal = {Nature},
  publisher = {Springer Science and Business Media LLC},
  author = {Kandala,  Abhinav and Mezzacapo,  Antonio and Temme,  Kristan and Takita,  Maika and Brink,  Markus and Chow,  Jerry M. and Gambetta,  Jay M.},
  year = {2017},
  month = sep,
  pages = {242–246}
}

@article{WeckerPRA2015,
  title = {Progress towards practical quantum variational algorithms},
  author = {Wecker, Dave and Hastings, Matthew B. and Troyer, Matthias},
  journal = {Phys. Rev. A},
  volume = {92},
  issue = {4},
  pages = {042303},
  numpages = {10},
  year = {2015},
  month = {Oct},
  publisher = {American Physical Society},
  doi = {10.1103/PhysRevA.92.042303},
  url = {https://link.aps.org/doi/10.1103/PhysRevA.92.042303}
}

@article{GrimsleyNatureComm2019,
  title = {An adaptive variational algorithm for exact molecular simulations on a quantum computer},
  volume = {10},
  ISSN = {2041-1723},
  url = {http://dx.doi.org/10.1038/s41467-019-10988-2},
  DOI = {10.1038/s41467-019-10988-2},
  number = {1},
  journal = {Nature Communications},
  publisher = {Springer Science and Business Media LLC},
  author = {Grimsley,  Harper R. and Economou,  Sophia E. and Barnes,  Edwin and Mayhall,  Nicholas J.},
  year = {2019},
  month = jul 
}

@article{GrimsleyNPJQI2022,
    author = "Grimsley, Harper R. and Barron, George S. and Barnes, Edwin and Economou, Sophia E. and Mayhall, Nicholas J.",
    title = "{Adaptive, problem-tailored variational quantum eigensolver mitigates rough parameter landscapes and barren plateaus}",
    eprint = "2204.07179",
    archivePrefix = "arXiv",
    primaryClass = "quant-ph",
    doi = "10.1038/s41534-023-00694-9",
    journal = "npj Quantum Inf.",
    volume = "9",
    number = "1",
    pages = "24",
    year = "2023"
}

@article{IbePhysRevR2022,
   title={Calculating transition amplitudes by variational quantum deflation},
   volume={4},
   ISSN={2643-1564},
   url={http://dx.doi.org/10.1103/PhysRevResearch.4.013173},
   DOI={10.1103/physrevresearch.4.013173},
   number={1},
   journal={Physical Review Research},
   publisher={American Physical Society (APS)},
   author={Ibe, Yohei and Nakagawa, Yuya O. and Earnest, Nathan and Yamamoto, Takahiro and Mitarai, Kosuke and Gao, Qi and Kobayashi, Takao},
   year={2022},
   month=mar}

@article{CiavarellaPRD2020,
   title={Algorithm for quantum computation of particle decays},
   volume={102},
   ISSN={2470-0029},
   url={http://dx.doi.org/10.1103/PhysRevD.102.094505},
   DOI={10.1103/physrevd.102.094505},
   number={9},
   journal={Physical Review D},
   publisher={American Physical Society (APS)},
   author={Ciavarella, Anthony},
   year={2020},
   month=nov }

@article{KitaevECCC1995,
  title={Quantum measurements and the Abelian Stabilizer Problem},
  author={Alexei Y. Kitaev},
  journal={Electron. Colloquium Comput. Complex.},
  year={1995},
  volume={TR96},
  url={https://api.semanticscholar.org/CorpusID:17023060}
}

@article{HiggottQuantum2019,
  title = {Variational Quantum Computation of Excited States},
  volume = {3},
  ISSN = {2521-327X},
  url = {http://dx.doi.org/10.22331/q-2019-07-01-156},
  DOI = {10.22331/q-2019-07-01-156},
  journal = {Quantum},
  publisher = {Verein zur Forderung des Open Access Publizierens in den Quantenwissenschaften},
  author = {Higgott,  Oscar and Wang,  Daochen and Brierley,  Stephen},
  year = {2019},
  month = jul,
  pages = {156}
}

@article{NakanishiPhysRevR2019,
   title={Subspace-search variational quantum eigensolver for excited states},
   volume={1},
   ISSN={2643-1564},
   url={http://dx.doi.org/10.1103/PhysRevResearch.1.033062},
   DOI={10.1103/physrevresearch.1.033062},
   number={3},
   journal={Physical Review Research},
   publisher={American Physical Society (APS)},
   author={Nakanishi, Ken M. and Mitarai, Kosuke and Fujii, Keisuke},
   year={2019},
   month=oct }

@misc{Sherbert2022,
      title={Orthogonal-ansatz VQE: Locating excited states without modifying a cost-function}, 
      author={Kyle Sherbert and Marco Buongiorno Nardelli},
      year={2022},
      eprint={2204.04361},
      archivePrefix={arXiv},
      primaryClass={quant-ph},
      url={https://arxiv.org/abs/2204.04361}, 
}

@article{WenQuantEng2021,
  title = {Variational quantum packaged deflation for arbitrary excited states},
  volume = {3},
  ISSN = {2577-0470},
  url = {http://dx.doi.org/10.1002/que2.80},
  DOI = {10.1002/que2.80},
  number = {4},
  journal = {Quantum Engineering},
  publisher = {Hindawi Limited},
  author = {Wen,  Jingwei and Lv,  Dingshun and Yung,  Man‐Hong and Long,  Gui‐Lu},
  year = {2021},
  month = sep 
}

@article{GochoNPJCompMat2023,
  title = {Excited state calculations using variational quantum eigensolver with spin-restricted ans\"{a}tze and automatically-adjusted constraints},
  volume = {9},
  ISSN = {2057-3960},
  url = {http://dx.doi.org/10.1038/s41524-023-00965-1},
  DOI = {10.1038/s41524-023-00965-1},
  number = {1},
  journal = {npj Computational Materials},
  publisher = {Springer Science and Business Media LLC},
  author = {Gocho,  Shigeki and Nakamura,  Hajime and Kanno,  Shu and Gao,  Qi and Kobayashi,  Takao and Inagaki,  Taichi and Hatanaka,  Miho},
  year = {2023},
  month = jan 
}

@article{GrimsleyQST2025,
doi = {10.1088/2058-9565/ad9fa2},
url = {https://dx.doi.org/10.1088/2058-9565/ad9fa2},
year = {2025},
month = {jan},
publisher = {IOP Publishing},
volume = {10},
number = {2},
pages = {025003},
author = {Grimsley, Harper R and Evangelista, Francesco A},
title = {Challenging excited states from adaptive quantum eigensolvers: subspace expansions vs. state-averaged strategies},
journal = {Quantum Science and Technology}
}

@misc{Chandani2024,
      title={Efficient charge-preserving excited state preparation with variational quantum algorithms}, 
      author={Zohim Chandani and Kazuki Ikeda and Zhong-Bo Kang and Dmitri E. Kharzeev and Alexander McCaskey and Andrea Palermo and C. R. Ramakrishnan and Pooja Rao and Ranjani G. Sundaram and Kwangmin Yu},
      year={2024},
      eprint={2410.14357},
      archivePrefix={arXiv},
      primaryClass={quant-ph},
      url={https://arxiv.org/abs/2410.14357}, 
}

@article{McCleanPRA2017,
  title = {Hybrid quantum-classical hierarchy for mitigation of decoherence and determination of excited states},
  author = {McClean, Jarrod R. and Kimchi-Schwartz, Mollie E. and Carter, Jonathan and de Jong, Wibe A.},
  journal = {Phys. Rev. A},
  volume = {95},
  issue = {4},
  pages = {042308},
  numpages = {10},
  year = {2017},
  month = {Apr},
  publisher = {American Physical Society},
  doi = {10.1103/PhysRevA.95.042308},
  url = {https://link.aps.org/doi/10.1103/PhysRevA.95.042308}
}

@misc{Parrish2019,
      title={Quantum Filter Diagonalization: Quantum Eigendecomposition without Full Quantum Phase Estimation}, 
      author={Robert M. Parrish and Peter L. McMahon},
      year={2019},
      eprint={1909.08925},
      archivePrefix={arXiv},
      primaryClass={quant-ph},
      url={https://arxiv.org/abs/1909.08925}, 
}

@article{StairJCTC2020,
author = {Stair, Nicholas H. and Huang, Renke and Evangelista, Francesco A.},
title = {A Multireference Quantum Krylov Algorithm for Strongly Correlated Electrons},
journal = {Journal of Chemical Theory and Computation},
volume = {16},
number = {4},
pages = {2236-2245},
year = {2020},
doi = {10.1021/acs.jctc.9b01125},
note ={PMID: 32091895},
URL = {https://doi.org/10.1021/acs.jctc.9b01125},
eprint = {https://doi.org/10.1021/acs.jctc.9b01125}
}

@article{CianciJCTC2024,
author = {Cianci, Cameron and Santos, Lea F. and Batista, Victor S.},
title = {Subspace-Search Quantum Imaginary Time Evolution for Excited State Computations},
journal = {Journal of Chemical Theory and Computation},
volume = {20},
number = {20},
pages = {8940-8947},
year = {2024},
doi = {10.1021/acs.jctc.4c00915},
    note ={PMID: 39352769},
URL = {https://doi.org/10.1021/acs.jctc.4c00915},
eprint = {https://doi.org/10.1021/acs.jctc.4c00915}
}

@misc{Zindorf2024,
      title={Efficient Implementation of Multi-Controlled Quantum Gates}, 
      author={Ben Zindorf and Sougato Bose},
      year={2024},
      eprint={2404.02279},
      archivePrefix={arXiv},
      primaryClass={quant-ph},
      url={https://arxiv.org/abs/2404.02279}, 
}

@misc{Kay2023,
      title={Tutorial on the Quantikz Package}, 
      author={Alastair Kay},
      year={2023},
      eprint={1809.03842},
      archivePrefix={arXiv},
      primaryClass={quant-ph},
      url={https://arxiv.org/abs/1809.03842}, 
}

@misc{Chen2023_1,
      title={Quantum Thermal State Preparation}, 
      author={Chi-Fang Chen and Michael J. Kastoryano and Fernando G. S. L. Brandão and András Gilyén},
      year={2023},
      eprint={2303.18224},
      archivePrefix={arXiv},
      primaryClass={quant-ph},
      url={https://arxiv.org/abs/2303.18224}, 
}

@misc{Chen2023_2,
      title={An efficient and exact noncommutative quantum Gibbs sampler}, 
      author={Chi-Fang Chen and Michael J. Kastoryano and András Gilyén},
      year={2023},
      eprint={2311.09207},
      archivePrefix={arXiv},
      primaryClass={quant-ph},
      url={https://arxiv.org/abs/2311.09207}, 
}

@inproceedings{BergamaschiIEEEFOCSProceedings2024,
   title={Quantum Computational Advantage with Constant-Temperature Gibbs Sampling},
   url={http://dx.doi.org/10.1109/FOCS61266.2024.00071},
   DOI={10.1109/focs61266.2024.00071},
   booktitle={2024 IEEE 65th Annual Symposium on Foundations of Computer Science (FOCS)},
   publisher={IEEE},
   author={Bergamaschi, Thiago and Chen, Chi-Fang and Liu, Yunchao},
   year={2024},
   month=oct, pages={1063–1085} }

@misc{Rajakumar2024,
      title={Gibbs Sampling gives Quantum Advantage at Constant Temperatures with O(1)-Local Hamiltonians}, 
      author={Joel Rajakumar and James D. Watson},
      year={2024},
      eprint={2408.01516},
      archivePrefix={arXiv},
      primaryClass={quant-ph},
      url={https://arxiv.org/abs/2408.01516}, 
}

@misc{Brunner2024,
      title={Lindblad engineering for quantum Gibbs state preparation under the eigenstate thermalization hypothesis}, 
      author={Eric Brunner and Luuk Coopmans and Gabriel Matos and Matthias Rosenkranz and Frederic Sauvage and Yuta Kikuchi},
      year={2024},
      eprint={2412.17706},
      archivePrefix={arXiv},
      primaryClass={quant-ph},
      url={https://arxiv.org/abs/2412.17706}, 
}

@article{VanDykePRR2024,
  title = {Scaling adaptive quantum simulation algorithms via operator pool tiling},
  author = {Van Dyke, John S. and Shirali, Karunya and Barron, George S. and Mayhall, Nicholas J. and Barnes, Edwin and Economou, Sophia E.},
  journal = {Phys. Rev. Res.},
  volume = {6},
  issue = {1},
  pages = {L012030},
  numpages = {8},
  year = {2024},
  month = {Feb},
  publisher = {American Physical Society},
  doi = {10.1103/PhysRevResearch.6.L012030},
  url = {https://link.aps.org/doi/10.1103/PhysRevResearch.6.L012030}
}

@article{FarrellPRXQ2024,
  title = {Scalable Circuits for Preparing Ground States on Digital Quantum Computers: The Schwinger Model Vacuum on 100 Qubits},
  author = {Farrell, Roland C. and Illa, Marc and Ciavarella, Anthony N. and Savage, Martin J.},
  journal = {PRX Quantum},
  volume = {5},
  issue = {2},
  pages = {020315},
  numpages = {32},
  year = {2024},
  month = {Apr},
  publisher = {American Physical Society},
  doi = {10.1103/PRXQuantum.5.020315},
  url = {https://link.aps.org/doi/10.1103/PRXQuantum.5.020315}
}

@inproceedings{SherbertIEEEConferenceProceeding2024,
    author = "Sherbert, Kyle and Furches, Jim and Shirali, Karunya and Economou, Sophia E. and Marrero, Carlos Ortiz",
    title = "{Adaptive Quantum Generative Training using an Unbounded Loss Function}",
    booktitle = "{2024 International Conference on Quantum Computing and Engineering}",
    eprint = "2408.00218",
    archivePrefix = "arXiv",
    primaryClass = "quant-ph",
    reportNumber = "DE-SC0012704, PNNL-SA-198421",
    doi = "10.1109/QCE60285.2024.00202",
    month = "7",
    year = "2024"
}

@Article{AsthanaRSCCS2023,
author ="Asthana, Ayush and Kumar, Ashutosh and Abraham, Vibin and Grimsley, Harper and Zhang, Yu and Cincio, Lukasz and Tretiak, Sergei and Dub, Pavel A. and Economou, Sophia E. and Barnes, Edwin and Mayhall, Nicholas J.",
title  ="Quantum self-consistent equation-of-motion method for computing molecular excitation energies{,} ionization potentials{,} and electron affinities on a quantum computer",
journal  ="Chem. Sci.",
year  ="2023",
volume  ="14",
issue  ="9",
pages  ="2405-2418",
publisher  ="The Royal Society of Chemistry",
doi  ="10.1039/D2SC05371C",
url  ="http://dx.doi.org/10.1039/D2SC05371C"}

@article{XieJCTC2022,
author = {Xie, Qing-Xing and Liu, Sheng and Zhao, Yan},
title = {Orthogonal State Reduction Variational Eigensolver for the Excited-State Calculations on Quantum Computers},
journal = {Journal of Chemical Theory and Computation},
volume = {18},
number = {6},
pages = {3737-3746},
year = {2022},
doi = {10.1021/acs.jctc.2c00159},
note ={PMID: 35621354},
URL = {https://doi.org/10.1021/acs.jctc.2c00159},
eprint = {https://doi.org/10.1021/acs.jctc.2c00159}
}

@article{PeruzzoNature2014,
title={A variational eigenvalue solver on a photonic quantum processor},
volume={5},
ISSN={2041-1723},
url={http://dx.doi.org/10.1038/ncomms5213},
DOI={10.1038/ncomms5213},
number={1},
journal={Nature Communications},
publisher={Springer Science and Business Media LLC},
author={Peruzzo, Alberto and McClean, Jarrod and Shadbolt, Peter and Yung, Man-Hong and Zhou, Xiao-Qi and Love, Peter J. and Aspuru-Guzik, Alán and O’Brien, Jeremy L.},
year={2014},
month=jul
}

@article{GardNPJ2020,
   title={Efficient symmetry-preserving state preparation circuits for the variational quantum eigensolver algorithm},
   volume={6},
   ISSN={2056-6387},
   url={http://dx.doi.org/10.1038/s41534-019-0240-1},
   DOI={10.1038/s41534-019-0240-1},
   number={1},
   journal={npj Quantum Information},
   publisher={Springer Science and Business Media LLC},
   author={Gard, Bryan T. and Zhu, Linghua and Barron, George S. and Mayhall, Nicholas J. and Economou, Sophia E. and Barnes, Edwin},
   year={2020},
   month=jan }

@article{BurtonPRR2024,
  title = {Accurate and gate-efficient quantum Ans\"atze for electronic states without adaptive optimization},
  author = {Burton, Hugh G. A.},
  journal = {Phys. Rev. Res.},
  volume = {6},
  issue = {2},
  pages = {023300},
  numpages = {17},
  year = {2024},
  month = {Jun},
  publisher = {American Physical Society},
  doi = {10.1103/PhysRevResearch.6.023300},
  url = {https://link.aps.org/doi/10.1103/PhysRevResearch.6.023300}
}

@article{TangPRXQ2021,
  title = {Qubit-ADAPT-VQE: An Adaptive Algorithm for Constructing Hardware-Efficient Ans\"atze on a Quantum Processor},
  author = {Tang, Ho Lun and Shkolnikov, V.O. and Barron, George S. and Grimsley, Harper R. and Mayhall, Nicholas J. and Barnes, Edwin and Economou, Sophia E.},
  journal = {PRX Quantum},
  volume = {2},
  issue = {2},
  pages = {020310},
  numpages = {16},
  year = {2021},
  month = {Apr},
  publisher = {American Physical Society},
  doi = {10.1103/PRXQuantum.2.020310},
  url = {https://link.aps.org/doi/10.1103/PRXQuantum.2.020310}
}

@misc{Ramoa2024,
      title={Reducing the Resources Required by ADAPT-VQE Using Coupled Exchange Operators and Improved Subroutines}, 
      author={Mafalda Ramôa and Panagiotis G. Anastasiou and Luis Paulo Santos and Nicholas J. Mayhall and Edwin Barnes and Sophia E. Economou},
      year={2024},
      eprint={2407.08696},
      archivePrefix={arXiv},
      primaryClass={quant-ph},
      url={https://arxiv.org/abs/2407.08696}, 
}

@article{AnastasiouPRR2024,
   title={TETRIS-ADAPT-VQE: An adaptive algorithm that yields shallower, denser circuit 
Ansätze},
   volume={6},
   ISSN={2643-1564},
   url={http://dx.doi.org/10.1103/PhysRevResearch.6.013254},
   DOI={10.1103/physrevresearch.6.013254},
   number={1},
   journal={Physical Review Research},
   publisher={American Physical Society (APS)},
   author={Anastasiou, Panagiotis G. and Chen, Yanzhu and Mayhall, Nicholas J. and Barnes, Edwin and Economou, Sophia E.},
   year={2024},
   month=mar
}

@article{FeniouSpringerCP2023,
   title={Overlap-ADAPT-VQE: practical quantum chemistry on quantum computers via overlap-guided compact Ansätze},
   volume={6},
   ISSN={2399-3650},
   url={http://dx.doi.org/10.1038/s42005-023-01312-y},
   DOI={10.1038/s42005-023-01312-y},
   number={1},
   journal={Communications Physics},
   publisher={Springer Science and Business Media LLC},
   author={Feniou, César and Hassan, Muhammad and Traoré, Diata and Giner, Emmanuel and Maday, Yvon and Piquemal, Jean-Philip},
   year={2023},
   month=jul 
}

@article{julia,
	title = {Julia: {A} {Fresh} {Approach} to {Numerical} {Computing}},
	volume = {59},
	issn = {0036-1445},
	shorttitle = {Julia},
	url = {https://epubs.siam.org/doi/10.1137/141000671},
	doi = {10.1137/141000671},
	number = {1},
	urldate = {2024-03-17},
	journal = {SIAM Review},
	author = {Bezanson, Jeff and Edelman, Alan and Karpinski, Stefan and Shah, Viral B.},
	month = jan,
	year = {2017},
	note = {Publisher: Society for Industrial and Applied Mathematics},
	pages = {65--98},
}

@article{optim,
  author  = {Mogensen, Patrick Kofod and Riseth, Asbj{\o}rn Nilsen},
  title   = {Optim: A mathematical optimization package for {Julia}},
  journal = {Journal of Open Source Software},
  year    = {2018},
  volume  = {3},
  number  = {24},
  pages   = {615},
  doi     = {10.21105/joss.00615}
}

@article{BetheZeitschriftPhysik1931,
	author = {Bethe, H. },
	date = {1931/03/01},
	doi = {10.1007/BF01341708},
	id = {Bethe1931},
	isbn = {0044-3328},
	journal = {Zeitschrift f{\"u}r Physik},
	number = {3},
	pages = {205--226},
	title = {Zur Theorie der Metalle},
	url = {https://doi.org/10.1007/BF01341708},
	volume = {71},
	year = {1931}}

@article{Xu:2023dgi,
    author = "Xu, G. and Guo, Y. B. and Li, X. and Wang, K. and Fan, Z. and Zhou, Z. S. and Liao, H. J. and Xiang, T.",
    title = "{Concurrent quantum eigensolver for multiple low-energy eigenstates}",
    doi = "10.1103/PhysRevA.107.052423",
    journal = "Phys. Rev. A",
    volume = "107",
    number = "5",
    pages = "052423",
    year = "2023"
}

@article{HendekovicChemPhysLett1982,
title = {On the energy variation method},
journal = {Chemical Physics Letters},
volume = {90},
number = {3},
pages = {198-201},
year = {1982},
issn = {0009-2614},
doi = {https://doi.org/10.1016/0009-2614(82)80024-9},
url = {https://www.sciencedirect.com/science/article/pii/0009261482800249},
author = {Josip Hendeković}
}

@article{EzzellQST2023,
doi = {10.1088/2058-9565/acc4e3},
url = {https://dx.doi.org/10.1088/2058-9565/acc4e3},
year = {2023},
month = {apr},
publisher = {IOP Publishing},
volume = {8},
number = {3},
pages = {035001},
author = {Ezzell, Nic and Ball, Elliott M and Siddiqui, Aliza U and Wilde, Mark M and Sornborger, Andrew T and Coles, Patrick J and Holmes, Zoë},
title = {Quantum mixed state compiling},
journal = {Quantum Science and Technology}
}

@article{ChenPRA2025,
  title = {Slack-variable approach for variational quantum semidefinite programming},
  author = {Chen, Jingxuan and Westerheim, Hanna and Holmes, Zo\"e and Luo, Ivy and Nuradha, Theshani and Patel, Dhrumil and Rethinasamy, Soorya and Wang, Kathie and Wilde, Mark M.},
  journal = {Phys. Rev. A},
  volume = {112},
  issue = {2},
  pages = {022607},
  numpages = {52},
  year = {2025},
  month = {Aug},
  publisher = {American Physical Society},
  doi = {10.1103/lwxq-4myj},
  url = {https://link.aps.org/doi/10.1103/lwxq-4myj}
}

@misc{Bittel2022,
      title={Fast gradient estimation for variational quantum algorithms}, 
      author={Lennart Bittel and Jens Watty and Martin Kliesch},
      year={2022},
      eprint={2210.06484},
      archivePrefix={arXiv},
      primaryClass={quant-ph},
      url={https://arxiv.org/abs/2210.06484}, 
}

@article{MitaraiPRA2018,
   title={Quantum circuit learning},
   volume={98},
   ISSN={2469-9934},
   url={http://dx.doi.org/10.1103/PhysRevA.98.032309},
   DOI={10.1103/physreva.98.032309},
   number={3},
   journal={Physical Review A},
   publisher={American Physical Society (APS)},
   author={Mitarai, K. and Negoro, M. and Kitagawa, M. and Fujii, K.},
   year={2018},
   month=sep }

%%%%%%%%%%%%%%%%%%%%%%%%%%%%%%%%%%%%%%%%%%%%%%%%%%%%%%%%%%%%%%%%%%%%%%%%%%%%%%%%%%%%%%%%%%
%%%%%%%%%%%%%%%%%%%%%%%%%%%%%%%%%%%%%%%%%%%%%%%%%%%%%%%%%%%%%%%%%%%%%%%%%%%%%%%%%%%%%%%%%%

\appendix
\clearpage

\section{Theorem 1}\label{app:Theorem1}

In this section, we will prove the following theorem:
\\\\
\textbf{Theorem 1:} The rank-$m$ state that minimizes the free energy $F=\langle H\rangle-\beta^{-1}S$ is the Gibbs state of the Hamiltonian, $H$, projected to the subspace spanned by its lowest $m$ eigenstates, $\{\ket{\psi_j}\}_{j=1}^m$:
\begin{equation}
    \rho_G^{(m)} = \sum_{j=1}^m\frac{e^{-\beta\,E_j}}{Z_m}\ket{\psi_j}\bra{\psi_j},
\end{equation}
where $\{E_j\}_{j=1}^m$ are the lowest eigenenergies of $H$, and $Z_m=\sum_{k=1}^m e^{-\beta\,E_k}$ is the truncated partition function. 
\\\\
\textit{Proof:} Consider an arbitrary rank-$m$ state, $\rho^{\mathcal{S}}$, restricted to a subspace $\mathcal{S}$. Let the Hamiltonian restricted to $\mathcal{S}$ be $H_{\mathcal{S}}$, and its corresponding Gibbs state be $\rho_{G}^{\mathcal{S}}$. The relative entropy/ Kullback-Leibler (KL) divergence between these two states is
\begin{align}\label{eq:KLDivergence}
    \mathcal{D}\left(\rho^{\mathcal{S}}\vert\vert\rho_G^{\mathcal{S}}\right)&=\Tr\left(\rho^{\mathcal{S}}\log\rho^{\mathcal{S}}\right)-\Tr\left(\rho^{\mathcal{S}}\log\rho_G^{\mathcal{S}}\right),\\
    &= -S(\rho^{\mathcal{S}})+\beta\Tr(\rho^{\mathcal{S}}H_{\mathcal{S}})+\log Z_{\mathcal{S}}.
\end{align}
Consider the free energy of $\rho^{\mathcal{S}}$:
\begin{equation}
    F(\rho^{\mathcal{S}})=\Tr(\rho^{\mathcal{S}}H)-\beta^{-1}S(\rho^{\mathcal{S}}).
\end{equation}
We can use Eq.~\eqref{eq:KLDivergence} to relate $F(\rho^{\mathcal{S}})$ to the KL divergence:
\begin{equation}\label{eq:FEnergySubspace}
    F(\rho^{\mathcal{S}})=\beta^{-1}\mathcal{D}\left(\rho^{\mathcal{S}}\vert\vert\rho_G^{\mathcal{S}}\right)-\beta^{-1}\log Z_{\mathcal{S}}.
\end{equation}
The goal is to minimize $F(\rho^{\mathcal{S}})$. The KL divergence is positive semi-definite, and will be zero iff $\rho^{\mathcal{S}}=\rho_G^{\mathcal{S}}$. Thus to minimize the free energy in Eq.~\eqref{eq:FEnergySubspace}, we just need to maximize the partition function $Z_{\mathcal{S}}$. 
\\\\
Let $\{\phi_k\}_{k=1}^m$ be a basis that spans $\mathcal{S}$. The partition function corresponding to $H_{\mathcal{S}}$ is
\begin{equation}
    Z_{\mathcal{S}}=\sum_{j=1}^m\exp\left(-\beta\bra{\phi_j}H\ket{\phi_j}\right),
\end{equation}
which is maximized when $\ket{\phi_j}=\ket{\psi_j}\,\,\forall\,j=1,\cdots,m$. In other words, the rank-$m$ subspace $\mathcal{S}$ that contains the Gibbs state that minimizes the free energy is the one spanned by the lowest $m$ eigenstates of the Hamiltonian $H$. 

This concludes the proof.

%%%%%%%%%%%%%%%%%%%%%%%%%%%%%%%%%%%%%%%%%%%%%%%%%%%%%%%%%%%%%%%%%%%%%%%%%%%%%%%%%%%%%%%%%%
%%%%%%%%%%%%%%%%%%%%%%%%%%%%%%%%%%%%%%%%%%%%%%%%%%%%%%%%%%%%%%%%%%%%%%%%%%%%%%%%%%%%%%%%%%

\section{Explicit form of $U_m$}\label{app:ExplicitUm}

In principle, there are several ways to construct $U_m$, using, for example, the quantum channels formalism. The evolution of the system density matrix can be treated as a quantum channel characterized by a set of Kraus operators. One could then find a unitary realization of the channel using ancillary qubits. In this subsection, we provide an intuitive parametrization in terms of a series of Givens rotations (see Eqs.~\eqref{eq:GivensExpMap},~\eqref{eq:GivensGens}) on the ancillary register, followed by CNOT gates that connect the ancillary and system registers. Finally, we have a permutation matrix on the system register that depends on the choice of the computational subspace. Note that this likely is not the most efficient way to implement $U_m$, but serves as an example. For this form of $U_m$, we require a minimal number of ancillary qubits $N_a=\lceil{\log_2m}\rceil$.

First, we write down a re-parametrization of $\vec{\mu}$ in terms of angles, using the $m$-dimensional spherical polar coordinates
\begin{equation}\label{eq:PolarParam}
    \sqrt{\mu_j}=
    \begin{cases}
        \left(\prod_{k=1}^{j-1}\sin(\varphi_k)\right)\cos(\varphi_j) & 1\leq j<m \\\\
        \left(\prod_{k=1}^{m-1}\sin(\varphi_k)\right) & j=m
    \end{cases},
\end{equation}
where $m$ is the number of parameters, corresponding to the number of non-zero eigenvalues in $\rho_m$. This re-parametrization maps the $m$ parameters $\vec{\mu}$, with the unit trace constraint to $m-1$ independent angles $\vec{\varphi}$. It also ensures the positivity of $\vec{\mu}$.

Next, we prepare on the ancillary register the pure state
\begin{equation}\label{eq:PostGivensState}
    \ket{\xi}=\sum_{k=1}^m\sqrt{\mu_k}\ket{k-1},
\end{equation}
where $\{\ket{k}\}$ are the computational basis elements. For instance, $\ket{3}$ on a 4-qubit ancillary register corresponds to $\ket{0011}$. This is done using a series of $m-1$ Givens rotations
\begin{equation}
    \ket{\xi}=\prod_{k=1}^{m-1}G^{(k,k+1)}(\varphi_k)\ket{0}^{\otimes N_a},
\end{equation}
where $G^{(k,k+1)}(\varphi_k)$ is the Givens rotation generated by the exponential map
\begin{equation}\label{eq:GivensExpMap}
    G^{(k,k+1)}(\varphi_k)\equiv \exp\left({i\,\varphi_k\gamma^{(k,k+1)}}\right),
\end{equation}
where the matrix elements of $\gamma^{(k,k+1)}$ are
\begin{equation}\label{eq:GivensGens}
    \left[\gamma^{(k,k+1)}\right]_{x,y}=i\left(\delta_{x,k}\,\delta_{y,k+1}-\delta_{x,k+1}\,\delta_{y,k}\right),
\end{equation}
using $\delta_{i,j}$ for the Kronecker delta. Geometrically, this corresponds to an $SO(2^{\lceil\log_2m\rceil})$ rotation from 
\begin{equation}
    \left(1,0,\cdots,0\right)\to\left(\sqrt{\mu_1},\cdots,\sqrt{\mu_m},\cdots\right),
\end{equation}
where the vectors are in the computational basis. The last set of $\cdots$ on the second vector corresponds to a padding with $2^{\lceil\log_2m\rceil}-m$ zeros.

%%%%%%%%%%%%%%%%%%%%%%%%%%%%%%%%%%%%%%%%%%%%%%%%%%%%%%%%%%%%%%%%%%%%%%%%%%%%%%%%%%%%%%%%%%

\subsection{Intuitive implementation}

\begin{figure}[ht!]
    \centering
    \begin{adjustbox}{width=\linewidth}
    \begin{quantikz}[classical gap=0.07cm]
    \lstick{$s_1$} & & & & & \gate[3]{P\left(\left\{\ket{c_k}\right\}\right)}& \rstick[3]{$\rho_3$}\\
    \lstick{$s_2$} & & & & \targ{} & &\\
    \lstick{$s_3$} & & & \targ{} & & &\\
    \lstick{$a_1$} & \gate[2]{G^{(1,2)}(\varphi_1)} &\gate[2]{G^{(2,3)}(\varphi_2)} & & \ctrl{-2} & &\\
    \lstick{$a_2$} & & & \ctrl{-2} & & &
    \end{quantikz}
    \end{adjustbox}
    \caption{An explicit circuit for $U_3$ with $N_s=3$. The reduced density matrix of the system register is $\rho_3$ in Eq.~\eqref{eq:rho_m} with $m=3$.}
    \label{fig:GivensAnsatz}
\end{figure}
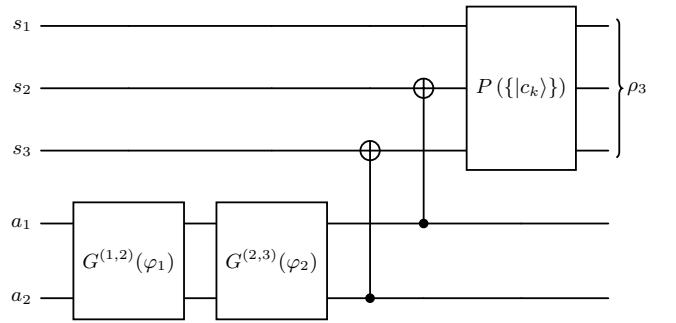

Following this, we perform a set of CNOT operations with the $j^{th}$ ancillary qubit as the control, and the $(N_s-m+j)^{th}$ system qubit as the target, for every ancillary qubit. Note that this is a \textit{single} layer of CNOT gates. The state on the system + ancillary register after this operation is
\begin{equation}\label{eq:GivCxState}
    \ket{\Omega(\vec{\varphi})}=\sum_{k=1}^m\sqrt{\mu_k}(\vec{\varphi})\left(\ket{k-1}\otimes\ket{k-1}\right),
\end{equation}
where the ordering of the tensor product is (system $\otimes$ ancilla). The next step is the application of a permutation matrix on the system register. These are a class of matrices that permute the computational basis. The particular permutation matrix $P\left(\left\{\ket{c_k}\right\}\right)$ we need is the one that maps the first $m$ computational basis elements $\left\{\ket{k-1}\right\}_{k=1}^{m}$ to the chosen computational subspace $\left\{\ket{c_k}\right\}_{k=1}^m$ giving us the state
\begin{equation}
    \ket{\Phi(\vec{\varphi})}=\sum_{k=1}^m\sqrt{\mu_k}(\vec{\varphi})\left(\ket{c_k}\otimes\ket{k-1}\right),
\end{equation}
Then, upon tracing out the ancillary qubits, we obtain
\begin{equation}
    \Tr_a\ket{\Phi(\vec{\varphi})}\bra{\Phi(\vec{\varphi})}=\sum_{k=1}^m\mu_k(\vec{\varphi})\ket{c_k}\bra{c_k}\equiv\rho_m
\end{equation}
on the system register, as needed. An illustrative example of the circuit of $U_3$ using the above construction for a system with three qubits is shown in Fig.~\ref{fig:GivensAnsatz}. Upon tracing out the ancillary qubits, one obtains the desired $\rho_3$ on the system register, as indicated.

An important consideration for larger $N_s$ is choosing a family of permutation matrices that can be implemented efficiently. We use this construction as an intuitive example. We could instead use $m$ multi-controlled gates with the ancillary register as the controls and the system register as the targets to map to the desired $N_s$ qubit computational basis subspace. We discuss this realization along with its scalability below.

%%%%%%%%%%%%%%%%%%%%%%%%%%%%%%%%%%%%%%%%%%%%%%%%%%%%%%%%%%%%%%%%%%%%%%%%%%%%%%%%%%%%%%%%%%

\subsection{Scalable Implementation}
\begin{figure}[ht!]
    \centering
    \begin{adjustbox}{width=\linewidth}
    \begin{quantikz}[classical gap=0.07cm]
    \lstick{$s_1$} & &
                    & & & \targ{} & \targ{} &\rstick[3]{$\rho_3$}\\
    \lstick{$s_2$} & &
                    & \targ{} & \targ{} & & \targ{} &\\
    \lstick{$s_3$}  & &
                    & & \targ{} & \targ{} & \targ{} &\\
    \lstick{$a_1$} & \gate[2]{G^{(1,2)}(\varphi_1)} &\gate[2]{G^{(2,3)}(\varphi_2)}
                    & \octrl{-2} & \octrl{-2} & \ctrl{-3} & \ctrl{-3}&\\
    \lstick{$a_2$} & &
                    & \octrl{-1} & \ctrl{-1} & \octrl{-1} & \ctrl{-1} &
    \end{quantikz}
    \end{adjustbox}
    \caption{An explicit circuit for $U_3$ with $N_s=3$, $\qty{c_k} = \qty{010,011,101,111}$. The reduced density matrix of the system register is $\rho_3$ in Eq.~\eqref{eq:rho_m} with $m=3$. The open circle denotes a control operation conditional on $\ket{0}$ rather than $\ket{1}$.}
    \label{fig:GivensAnsatzSansPermutation}
\end{figure}
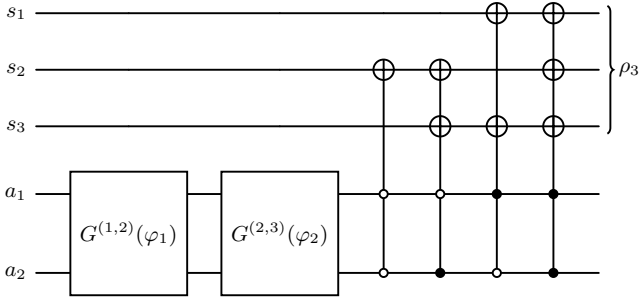

After applying the Givens rotations to prepare the state $\ket{\xi}$ in Eq.~\eqref{eq:PostGivensState} on the ancillary register, we perform a series of $m$ multi-controlled operations to prepare each chosen computational subspace element $\ket{c_k}$, conditional on the ancillary register with the value $\ket{k-1}$.
The state on the system and ancillary registers after this operation is
\begin{equation}
\ket{\Phi(\vec{\varphi})}=\sum_{k=1}^m\sqrt{\mu_k}(\vec{\varphi})\left(\ket{c_k}\otimes\ket{k-1}\right),
\end{equation}
where the ordering of the tensor product is (system $\otimes$ ancilla). 
Then, upon tracing out the ancillary qubits, we obtain
\begin{equation}
    \Tr_a\ket{\Phi(\vec{\varphi})}\bra{\Phi(\vec{\varphi})}=\sum_{k=1}^m\mu_k(\vec{\varphi})\ket{c_k}\bra{c_k}\equiv\rho_m
\end{equation}
on the system register, as needed. An illustrative example of the circuit of $U_3$ using the above construction for a system with three qubits is shown in Fig.~\ref{fig:GivensAnsatzSansPermutation}. Upon tracing out the ancillary qubits, one obtains the desired $\rho_3$ on the system register, as indicated.

Each of the $m-1$ Givens rotations can be implemented by a number of CNOT gates that scale linearly~\cite{Zindorf2024} in the number of qubits in the ancillary register $\mathcal{O}(\lceil\log_2 m\rceil)$. Similarly, each of the $m$ multi-controlled circuits to prepare each $\ket{c_k}$ can be decomposed into $O(m N_s)$ CNOT gates. Therefore, the total depth of $U_m$ scales as $O(m\,\lceil\log_2m\rceil N_s)$.

%%%%%%%%%%%%%%%%%%%%%%%%%%%%%%%%%%%%%%%%%%%%%%%%%%%%%%%%%%%%%%%%%%%%%%%%%%%%%%%%%%%%%%%%%%
%%%%%%%%%%%%%%%%%%%%%%%%%%%%%%%%%%%%%%%%%%%%%%%%%%%%%%%%%%%%%%%%%%%%%%%%%%%%%%%%%%%%%%%%%%

\section{Ancilla-free implementation}\label{app:AncillaFree}
In this section, we will propose a way to train the basis-change unitary $V_A(\vec{\theta}^*)$ without the use of ancillary qubits. This approach stems from the interpretation of a mixed state as a convex sum of pure states:
\begin{equation}
    \rho=\sum_{i}p_i\ket{\phi_i}\bra{\phi_i},
\end{equation}
instead of as having partial information of a larger pure state. In this picture, measuring an observable on a mixed state corresponds to an ensemble average over the pure states making up the convex sum:
\begin{equation}
    \langle\mathcal{O}\rangle\equiv\Tr(\rho\,\mathcal{O})=\sum_{i}p_i\bra{\phi_i}\mathcal{O}\ket{\phi_i}.
\end{equation}

The method prescribed in the main text involves preparing a classical (diagonal in the computational basis) density matrix $\rho_m$ using a purification, and then changing the basis using $V_A(\vec{\theta})$. The reduced-rank approximation enabled the use of a lower number of $\lceil\log_2m\rceil$ ancillary qubits. The cost function and gradients for ADAPT-VQE are measured following this, to train $V_A(\vec{\theta})$ to prepare the Gibbs state. 

An alternative method to effectively prepare
\begin{equation}
    \rho_m=\sum_{k=1}^m\mu_k\ket{c_k}\bra{c_k}
\end{equation}
is to sample $\ket{c_k}$ with probability $\mu_k$ and then evolve it under $V_A(\vec \theta)$. Using this state we can measure the cost function and its gradients. This approach prepares the same density matrix $\rho_m$ but as a statistical mixture rather than as the reduced state of a larger entangled state. This approach can also make the classical simulator more efficient.  If we use state-vector simulations, then the method we described earlier required memory $\mathcal{O}(m 2^{N_s})$, and a similar number of steps for each matrix-vector product.  The classical simulator could also sample from $\vec\mu$ and then use only $\mathcal{O}(2^{N_s})$ memory, albeit with the cost of introducing some variance into its estimates, similar to the variance already faced by the quantum algorithm.

The gradients of the free energy are given by
\begin{align}
\begin{split}
    \frac{\partial F}{\partial\mu_j}&=\frac{\partial}{\partial\mu_j}\langle H\rangle(\vec{\mu},\vec{\theta})-\beta^{-1}\frac{\partial}{\partial\mu_j}S(\vec{\mu}),\\
    &=\bra{c_j}V_A^{\dagger}(\vec{\theta})H V_A(\vec{\theta})\ket{c_j}+\beta^{-1}(\log\mu_j+1),
\end{split}
\end{align}
\begin{align}
\begin{split}
    \frac{\partial F}{\partial\theta_j}&=\frac{\partial}{\partial\theta_j}\langle H\rangle(\vec{\mu},\vec{\theta}),\\
    %&=\frac{2}{n_{\text{S}}}\Re\left\{\sum_{\sigma=1}^{n_{\text{S}}}\bra{c_{\sigma}}V_A^{\dagger}(\vec{\theta})H\frac{\partial}{\partial\theta_j}V_A(\vec{\theta})\ket{c_{\sigma}}\right\},\\
    &= 2\Re\left\{\sum_{k=1}^{m}\mu_k\bra{c_k}V_A^{\dagger}(\vec{\theta})H\frac{\partial}{\partial\theta_j}V_A(\vec{\theta})\ket{c_k}\right\}.
\end{split}
\end{align}
%where the arrow again indicates the limit of infinite samples, and 
The derivative in the last two lines is given by 
\begin{equation}
    \frac{\partial}{\partial\theta_j}V_A(\vec{\theta})=i\prod_{p=N_i}^je^{i\,\theta_p\,T_p}\,T_j\prod_{p=j-1}^1\,e^{i\,\theta_p\,T_p},
\end{equation}
where $N_i$ is the length of the ansatz after $i$ adaptive iterations.
This ancilla-free implementation of TEPID-ADAPT is an alternative to preparing a reduced density matrix of a purification and measuring the required observables. At low-temperatures, this amounts to classically sampling from a small number of computational basis states using the probability distribution $\{\mu_k\}_{k=1}^m$, which can be done efficiently. This can be beneficial in the near term as it leads to shallower circuits, which are less affected by device noise.

%%%%%%%%%%%%%%%%%%%%%%%%%%%%%%%%%%%%%%%%%%%%%%%%%%%%%%%%%%%%%%%%%%%%%%%%%%%%%%%%%%%%%%%%%%
%%%%%%%%%%%%%%%%%%%%%%%%%%%%%%%%%%%%%%%%%%%%%%%%%%%%%%%%%%%%%%%%%%%%%%%%%%%%%%%%%%%%%%%%%%

\section{Analytical gradients}\label{App:AnalyticalGradients}
Analytical gradients of the parametrized cost function as defined in Eq.~\eqref{eq:FEnergtparam}, rewritten here for convenience
\begin{equation}
    F(\vec{\mu},\vec{\theta})=\langle H\rangle(\vec{\mu},\vec{\theta})-\beta^{-1}S(\vec{\mu}),
\end{equation} 
can be used to make the classical optimization and the quantum measurement of the gradients more efficient. In this appendix, we will write down these analytical gradients. Using the explicit parametrization of the $\left\{\mu_j\right\}$ in Eq.~\eqref{eq:PolarParam}, we can write down the exact gradients for the free energy $F(\vec{\varphi},\vec{\theta})$ 
\begin{align}
    \frac{\partial F}{\partial\varphi_j}&=\frac{\partial}{\partial\varphi_j}\langle H\rangle(\vec{\varphi},\vec{\theta})-\beta^{-1}\frac{\partial}{\partial\varphi_j}S(\vec{\varphi}),\nonumber\\\nonumber\\
    \frac{\partial F}{\partial\theta_j} &=\frac{\partial}{\partial\theta_j}\langle H\rangle(\vec{\varphi},\vec{\theta}).
\end{align}
Let us start with the entropy, which purely depends on $\vec{\varphi}$ :
\begin{equation}
    S(\vec{\varphi})=-\sum_k\mu_k(\vec{\varphi})\log\mu_k(\vec{\varphi}).
\end{equation}
The analytical partial derivatives are 
\begin{equation}
    \frac{\partial S}{\partial \varphi_j}=-2\sum_{k=1}^m\sqrt{\mu_k}\left(\log\mu_k+1\right)\frac{\partial\sqrt{\mu_k}}{\partial\varphi_j},
\end{equation}
where the Jacobian of the variable transformation in Eq.~\ref{eq:PolarParam} is given by
\begin{equation}\label{eq:Jacobian}
    \frac{\partial\sqrt{\mu_j}}{\partial\varphi_l}=
    \begin{cases}
        \left(\prod\limits_{k=1}^{l-1}\sin(\varphi_k)\right)\cos(\varphi_l)& j<m,l<j \\
        \hspace{1.7cm}\times\left(\prod\limits_{k=l+1}^{j-1}\cos(\varphi_j)\right) \\\\
        -\left(\prod\limits_{k=1}^{m-1}\sin(\varphi_k)\right) & j<m, l=j \\\\
        0 & j<m, l>j \\\\
        \left(\prod\limits_{k=1}^{l-1}\sin(\varphi_k)\right)\cos(\varphi_l)& j=m\\\hspace{1.7cm}\times\left(\prod\limits_{k=l+1}^{j}\sin(\varphi_k)\right).
    \end{cases}.
\end{equation}
Next, we compute the analytical gradients for the energy term $\langle H\rangle(\vec{\varphi},\vec{\theta})$. First, let us consider the state evolved after $N_i$ operators have been adaptively added to the ansatz
\begin{equation}
    \ket{\Psi(\vec{\varphi},\vec{\theta})}=(V_A(\vec{\theta})\otimes\mathbb{I})\ket{\Phi(\vec{\varphi})},
\end{equation}
where $\ket{\Phi}$ is the state defined in Eq.~\eqref{eq:GivCxState}, and $V_A(\vec{\theta})$ is the adaptively generated unitary (see Fig.~\ref{fig:BlockAnsatz})
\begin{equation}
    V_A(\vec{\theta})\equiv\prod\limits_{k=N_i}^{1}e^{i\,\theta_k\,T_k},
\end{equation}
where $T_k$ is the $k^{th}$ operator in ansatz. The parametrized energy is then given by 
\begin{equation}
       \langle H\rangle(\vec{\varphi},\vec{\theta})=\bra{\Psi(\vec{\varphi},\vec{\theta})}(H\otimes\mathbb{I})\ket{\Psi(\vec{\varphi},\vec{\theta})},
\end{equation}
and its analytical partial derivatives are
\begin{align}
    \frac{\partial}{\partial\varphi_j}\langle H\rangle(\vec{\varphi},\vec{\theta})&=\left(\frac{\partial}{\partial\varphi_j}\bra{\Phi(\vec{\varphi})}\right)(V_A^{\dagger}(\vec{\theta})\,H\otimes\mathbb{I})\ket{\Psi(\vec{\varphi},\vec{\theta})}\nonumber\\&+\bra{\Psi(\vec{\varphi},\vec{\theta})}(H\,V_A(\vec{\theta})\otimes\mathbb{I})\left(\frac{\partial}{\partial\varphi_j}\ket{\Phi(\vec{\varphi})}\right),\nonumber\\\nonumber\\
    %%%%%%%%%%%%%%%%%%%%%%%%
    \frac{\partial}{\partial\theta_j}\langle H\rangle(\vec{\varphi},\vec{\theta}) &=\bra{\Phi(\vec{\varphi})}\left(\left(\frac{\partial}{\partial\theta_j}V_A^{\dagger}(\vec{\theta})\right)H\otimes\mathbb{I}\right)\ket{\Psi(\vec{\varphi},\vec{\theta})}\nonumber\\&+\bra{\Psi(\vec{\varphi},\vec{\theta})}\left(H\left(\frac{\partial}{\partial\theta_j}V_A(\vec{\theta})\right)\otimes\mathbb{I}\right)\ket{\Phi(\vec{\varphi})},
\end{align}
where
\begin{align}
    \frac{\partial}{\partial\varphi_j}\ket{\Phi(\vec{\varphi})}&=\sum_{j=1}^m\frac{\partial\sqrt{\mu_j}}{\partial\varphi_l}\ket{c_j}\otimes\ket{j-1},\nonumber\\\nonumber\\
    \frac{\partial}{\partial\theta_j}V_A(\vec{\theta})&=i\left(\prod\limits_{k=N_i}^{j}e^{i\,\theta_k\,T_k}\right)T_j\left(\prod\limits_{k=j-1}^{1}e^{i\,\theta_k\,T_k}\right).
\end{align}

These analytical expressions for the gradients enables their efficient measurement on a quantum computer, avoiding the need for finite differences. A parameter-shift rule~\cite{MitaraiPRA2018}:
\begin{multline}
    \frac{\partial C}{\partial\phi_j}=\frac{1}{2\sin\alpha}\left(\text{Tr}\left(O\,U^{\dagger}(\vec{\phi}_+)\,\rho\,U(\vec{\phi}_+)\right)\right.\\\left.-\left(O\,U^{\dagger}(\vec{\phi}_-)\,\rho\,U(\vec{\phi}_-)\right)\right),
\end{multline}
which entails evolving the state by two circuits that look like the ansatz with the relevant parameter shifted by a constant $\alpha$, can be used.

%%%%%%%%%%%%%%%%%%%%%%%%%%%%%%%%%%%%%%%%%%%%%%%%%%%%%%%%%%%%%%%%%%%%%%%%%%%%%%%%%%%%%%%%%%
%%%%%%%%%%%%%%%%%%%%%%%%%%%%%%%%%%%%%%%%%%%%%%%%%%%%%%%%%%%%%%%%%%%%%%%%%%%%%%%%%%%%%%%%%%

\section{Choice of initial state}\label{App:InitialState}

In this appendix we address the choice of initial parameters $\vec{\mu}$ for the initial state $\rho_m$ for our algorithm before ADAPT-VQE starts to build the basis-change unitary $V_A$. In our approach, we initialize the state preparation ansatz to prepare the state $\ket{\Phi(\vec{\varphi}_0)}$ in Eq.~\eqref{eq:GivCxState} on the extended system, where $\vec{\varphi}_0$ is the choice of the initial state preparation parameters. Before we adaptively build the rest of the ansatz out on the system register, we allow the algorithm to classically optimize the $\vec{\varphi}$ parameters using free energy as the cost function.

It is straightforward to motivate what the result of this optimization would be. By construction, $\rho_m$ is diagonal in the computational basis. As a result, the expectation value of any off-diagonal portion of the Hamiltonian on $\rho_m$ will vanish. So, the parametrized cost function for this first round of optimization is
\begin{equation}
    F = \Tr(\rho_m(\vec{\mu})\,H_D)+\beta^{-1}\sum_{k=1}^m\mu_k\log\mu_k,
\end{equation}
where $H_D$ is the diagonal portion of the Hamiltonian. In this case, a good choice of the initial $\vec{\mu}$ can be 
\begin{equation}
    \mu_k=\frac{1}{\tilde{Z}_m}e^{-\beta H_{c_k,c_k}} \text{ , with   } \tilde{Z}_m = \sum_{k=1}^m e^{-\beta H_{c_k,c_k}},    
\end{equation}
where $H_{i,j}$ is the (i,j)$^{th}$ element of the matrix representation of the Hamiltonian in the computational basis.

In the extreme case where the Hamiltonian is fully off-diagonal, the first term above vanishes, and the optimization would choose the state that maximizes the entropy term---the maximally mixed state in the truncated subspace:
\begin{equation}
    \rho_m\longrightarrow\frac{1}{m}\text{diag}(\underbrace{1,\cdots,1}_{m},0,\cdots,0),
\end{equation}
where for convenience we have chosen an ordering where the relevant computational subspace occupies the first $m$ entries. In cases where the Hamiltonian is mostly off-diagonal, one could skip the initial optimization step and instead opt to start with this choice of initial parameters $\vec{\mu}$ instead.

%%%%%%%%%%%%%%%%%%%%%%%%%%%%%%%%%%%%%%%%%%%%%%%%%%%%%%%%%%%%%%%%%%%%%%%%%%%%%%%%%%%%%%%%%%
%%%%%%%%%%%%%%%%%%%%%%%%%%%%%%%%%%%%%%%%%%%%%%%%%%%%%%%%%%%%%%%%%%%%%%%%%%%%%%%%%%%%%%%%%%

\section{Tolerance and convergence}\label{App:Tolerances}
\begin{figure}
    \centering
    \includegraphics[width=\linewidth]{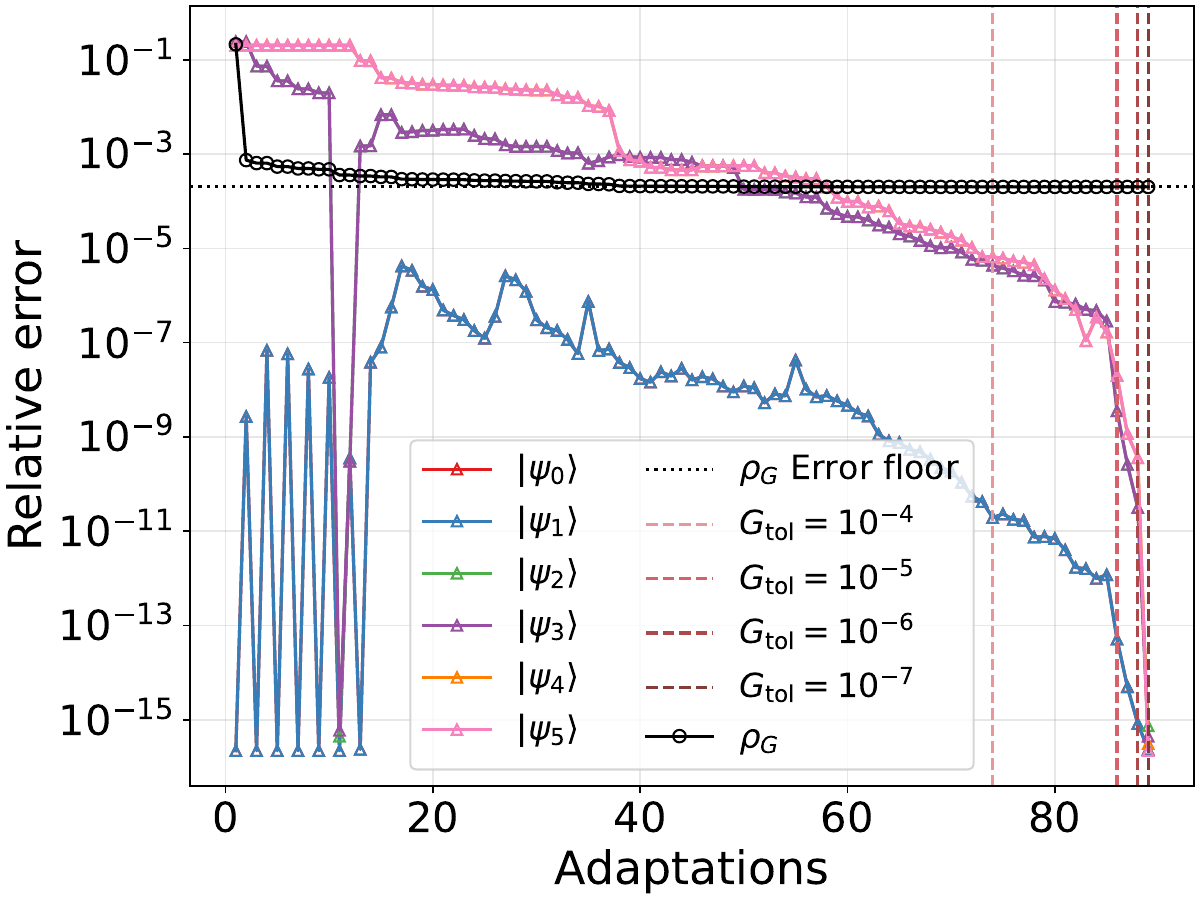}
    \caption{The relative (free) energy error as a function of the number of adaptations for the $\beta=3.0$ Gibbs state and eigenstates in the $m=6$ subspace for the Heisenberg model with $6$ spins for $J_z=-1.5$. The vertical dotted lines indicate convergence of TEPID-ADAPT for various pool gradient tolerances.}
    \label{fig:Gtolscan_Ferro}
\end{figure}

\begin{figure}
    \centering
    \includegraphics[width=\linewidth]{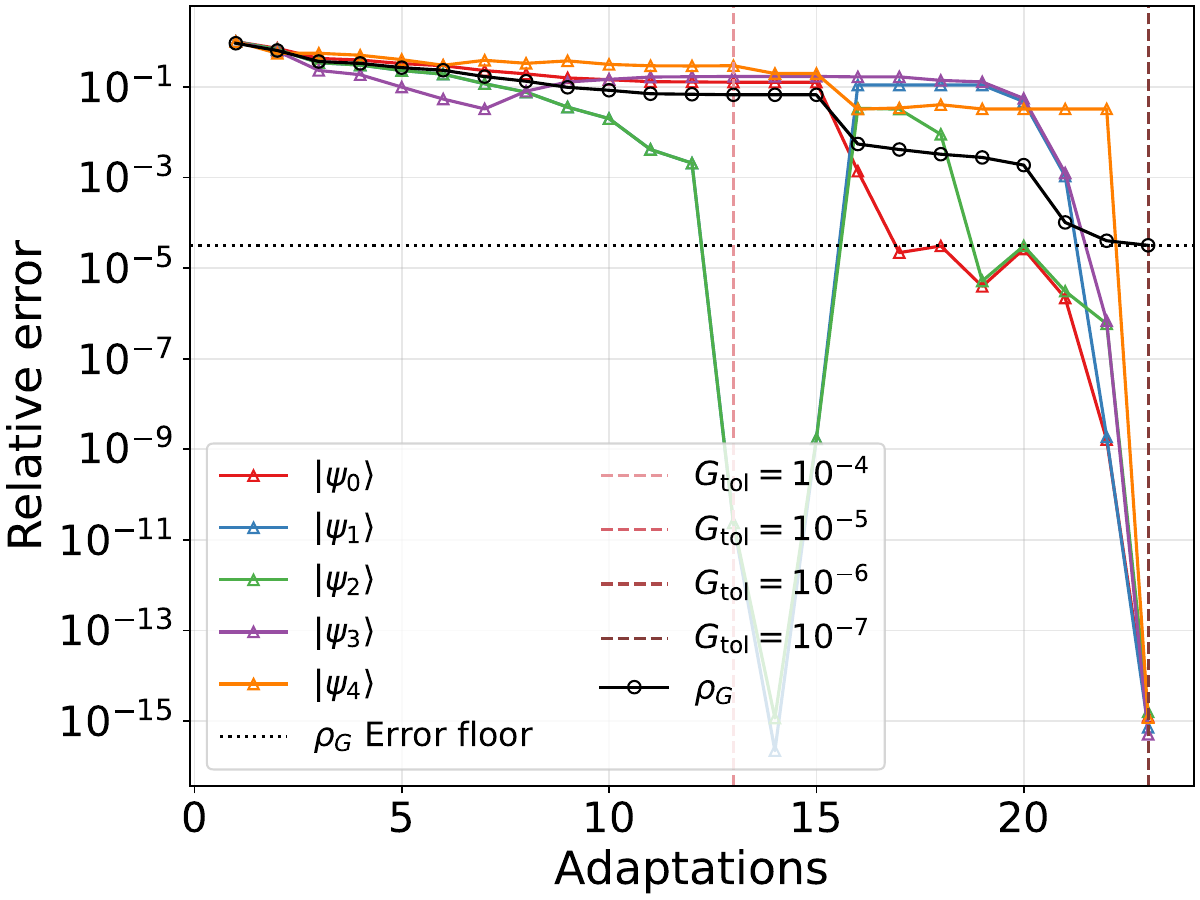}
    \caption{The relative (free) energy error as a function of the number of adaptations for the $\beta=3.0$ Gibbs state and eigenstates in the $m=5$ subspace for the Heisenberg model with $6$ spins for $J_z=0.0$. The vertical dotted lines indicate convergence of TEPID-ADAPT for various pool gradient tolerances.}
    \label{fig:Gtolscan_Para}
\end{figure}

\begin{figure}
    \centering
    \includegraphics[width=\linewidth]{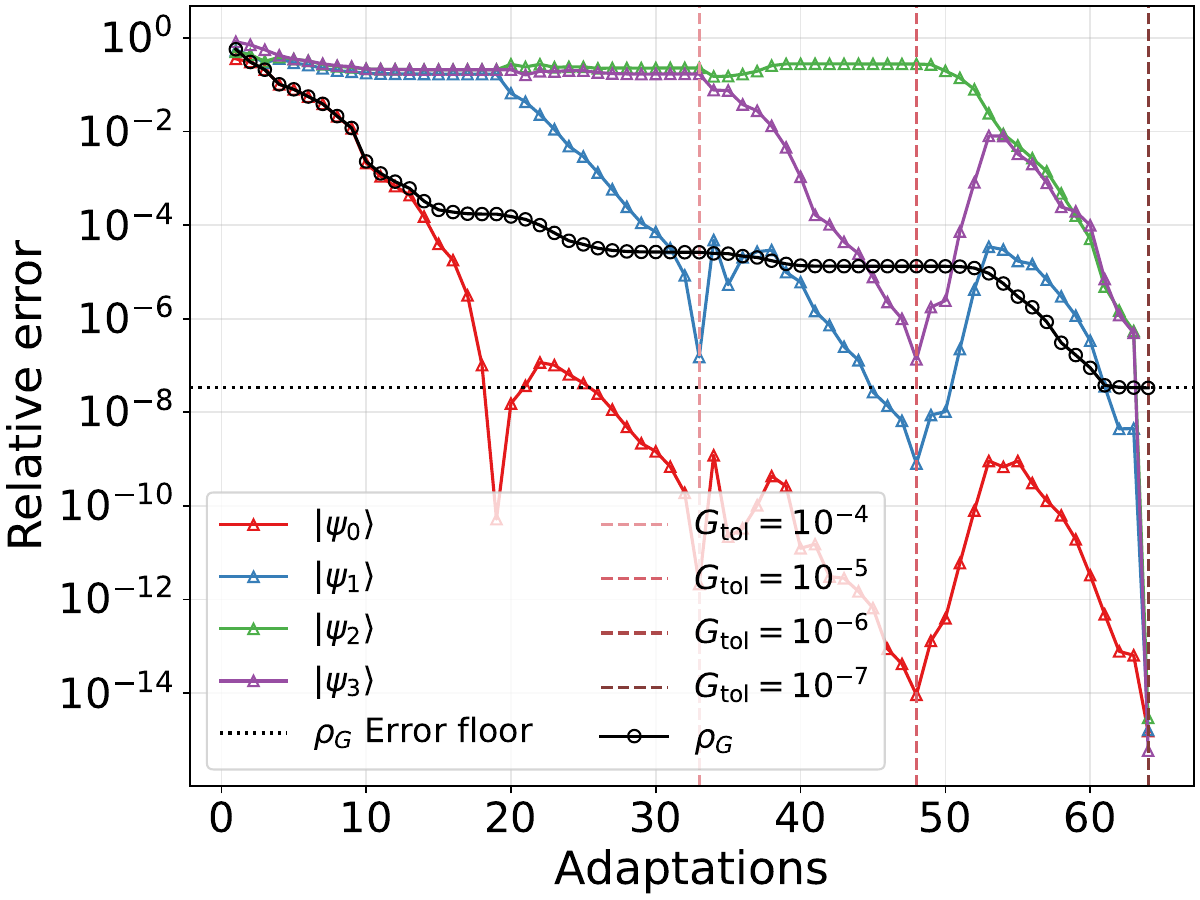}
    \caption{The relative (free) energy error as a function of the number of adaptations for the $\beta=3.0$ Gibbs state and eigenstates in the $m=4$ subspace for the Heisenberg model with $6$ spins for $J_z=1.5$. The vertical dotted lines indicate convergence of TEPID-ADAPT for various pool gradient tolerances.}
    \label{fig:Gtolscan_AFerro}
\end{figure}

In this Appendix, we study the effect of the pool gradient threshold on the convergence to the Gibbs state and low-energy eigenstates for various phases of the Heisenberg XXZ model. We use a BFGS scheme for the VQE subroutine of TEPID-ADAPT, implemented with the Julia programming language \cite{julia,optim}. In this work, we set the BFGS convergence criterion to be 
\begin{equation}\label{eq:BFGStol}
    \left\vert\left\vert\vec{\nabla}F(\vec{\mu},\vec{\theta})\right\vert\right\vert_{\infty}\leq 10^{-10},
\end{equation}
where $\vert\vert\boldsymbol{\cdot}\vert\vert_{\infty}$ is the infinity norm. Note that this tight convergence criterion is not a realistic one, but is used for the state-vector simulation results presented in this work to test the algorithm. We also have an ADAPT convergence criterion for the vector of local partial derivatives of the cost function for the pool operators
\begin{equation}\label{eq:Pooltol}
    \left\vert\left\vert\left\{\frac{\partial F}{\partial\theta_k}\right\}_{k=1}^{N_p}\right\vert\right\vert_{\infty}\leq \epsilon,
\end{equation}
where $N_p$ is the number of operators in the pool. In ADAPT-VQE, these local partial derivatives are used as a metric to choose the next operator to add to the ansatz. In the main text, for all plots, we set $\epsilon=10^{-6}$. In Figs.~\ref{fig:Gtolscan_Ferro},~\ref{fig:Gtolscan_Para},~\ref{fig:Gtolscan_AFerro}, we plot for a fixed cutoff $m$, the relative errors of free energy for the Gibbs state at a given temperature (black line) and of the energy for the eigenstates in the truncated eigenspace. 

We plot this as a function of the number of ``Adaptations'', or the number of parameters in the ansatz. Upon convergence of the VQE subroutine after the addition of the $n^{\text{th}}$ operator, we take the adaptive ansatz on the system register, $V^{(n)}_{A}$ and act on the chosen computational subspace $\left\{c_k\right\}_{k=1}^m$ to obtain the corresponding eigenstates $\left\{\psi_k\right\}_{k=1}^m$. We then measure the energy and compute the relative energy error with exact diagonalization. These are indicated by the different colors in Figs.~\ref{fig:Gtolscan_Ferro},~\ref{fig:Gtolscan_Para},~\ref{fig:Gtolscan_AFerro}. The vertical dashed lines in the plots show when both the convergence criteria Eqs.~\eqref{eq:BFGStol},~\eqref{eq:Pooltol} are met for different values of $\epsilon$, as shown in the plot legends. The black horizontal dashed line is the error floor of the free energy of the prepared Gibbs state. This is the closest one can get to the true Gibbs state given a particular truncation parameter $m$.

We see qualitative differences between the relative free energy curves (black lines) in the above figures. In the ferromagnetic phase Fig.~\ref{fig:Gtolscan_Ferro}, we see a sharp drop in the relative error with a small number of adaptations. This is likely a consequence of the fact that the computational basis states $\ket{000000},\ket{111111}$ are the exact ground states of the model. But after this initial drop, TEPID-ADAPT takes some time to find the other eigenstates. This causes the free energy to be relatively stagnant throughout the process. This feature is also a consequence of the relatively large energy gap between the ground and first excited eigenspaces.

In Figs.~\ref{fig:Gtolscan_Para},~\ref{fig:Gtolscan_AFerro}, we find a steeper free energy curve towards the end of TEPID-ADAPT for the higher tolerances. Given this behavior, it is instructive to try tighter values of the pool gradient tolerance $\epsilon$ to make sure ADAPT has converged to the correct state. 

%%%%%%%%%%%%%%%%%%%%%%%%%%%%%%%%%%%%%%%%%%%%%%%%%%%%%%%%%%%%%%%%%%%%%%%%%%%%%%%%%%%%%%%%%%
%%%%%%%%%%%%%%%%%%%%%%%%%%%%%%%%%%%%%%%%%%%%%%%%%%%%%%%%%%%%%%%%%%%%%%%%%%%%%%%%%%%%%%%%%%

\section{Scaling with $N_s$}\label{App:Scaling}
In this section, we numerically study how the truncation parameter $m$ (with fixed $\beta$) and the inverse temperature $\beta$ (with fixed $m$) scale with the system size $N_s$ for various phases of the Heisenberg XXZ model. More concretely, we find the minimum value of the truncation parameter $m_{\text{min}}$ for which the fidelity between the prepared and exact Gibbs states for a given temperature reaches a fixed threshold. Similarly, we also find the least inverse temperature $\beta_{\text{min}}$ with a fixed $m$ for which the fidelity with the exact Gibbs state reaches a fixed threshold. We use exact diagonalization for this study, using $2\leq N_s\leq 15$. We perform power-law and exponential fits to determine the scalings, where appropriate
\begin{align}\label{eq:FitModelsmvsN}
    f_P(N_s,a,b)&=a\,N_s^{b}, \nonumber\\
    f_E(N_s,a,b)&=a(e^{b\,N_s}-1),
\end{align}
where $a,b$ are the parameters we fit. We note that for gapped, locally interacting, translationally invariant systems, one expects $m_{\text{min}}$ to scale exponentially in the large $N_s$ limit. In this section, we numerically test this expectation for moderate values of $N_s$. It is unclear what the expected scaling would be for other types of models. In Figs.~\ref{fig:mvsN_Ferro},~\ref{fig:mvsN_Para},~\ref{fig:mvsN_AFerro}, we show the scaling of $m_{\text{min}}$ with $N_s$ for different fidelity thresholds, as indicated in the legends, for a fixed $\beta=3.0$. The solid black line corresponds to the $m=2^N$ curve, for reference. The power law and exponential fits in Eq.~\eqref{eq:FitModelsmvsN} are shown, respectively, using dotted and dashed lines. Figs.~\ref{fig:betavsN_Ferro},~\ref{fig:betavsN_Para},~\ref{fig:betavsN_AFerro} show the scaling of $\beta_{\text{min}}$ with $N_s$ for different fidelity thresholds for a fixed $m=4$. We use the mean squared error as a measure of the quality of the fits 
\begin{equation}
    \text{MSE}=\frac{1}{n_S}\sum_{k=1}^{n_S}\left(y_k-f(x_k)\right)^2,
\end{equation}
where $n_S$ is the number of samples, $\left\{\left(x_k,y_k\right)\right\}_{k=1}^{n_S}$ is the data set, and $f$ is the fitting model. If the MSE is zero, the model perfectly describes the data. If the MSE is large, the model is not a good fit for the data.

%%%%%%%%%%%%%%%%%%%%%%%%%%%%%%%%%%%%%%%%%%%%%%%%%%%%%%%%%%%%%%%%%%%%%%%%%%%%%%%%%%%%%%%%%%

\subsection{Scaling of $m_{\text{min}}$ with $N_s$}
\begin{figure}
    \centering
    \includegraphics[width=\linewidth]{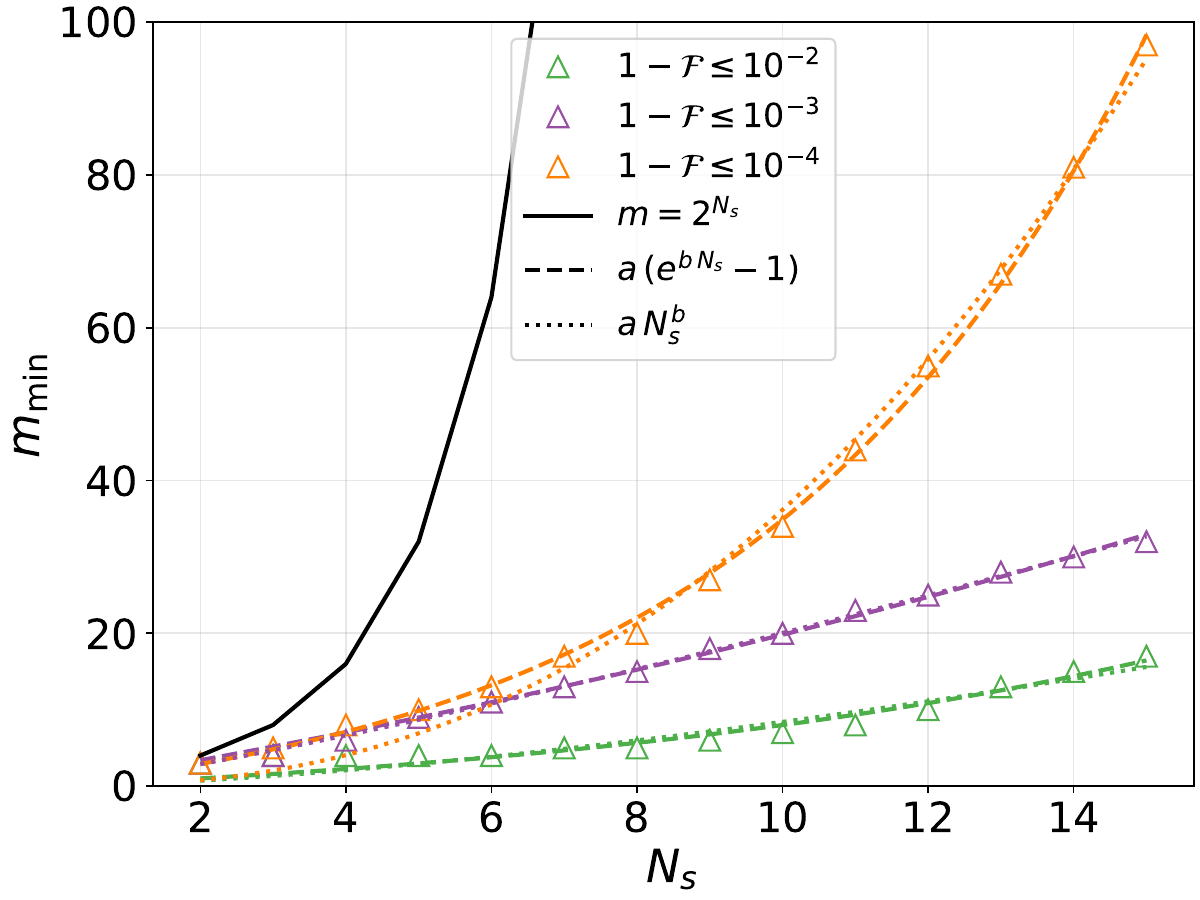}
    \caption{The scaling of $m_{\text{min}}$ required to achieve a given infidelity with $N_s$ in the ferromagnetic phase. The colors indicate different infidelity thresholds, as shown in the legend.}
    \label{fig:mvsN_Ferro}
\end{figure}

\begin{figure}
    \centering
    \includegraphics[width=\linewidth]{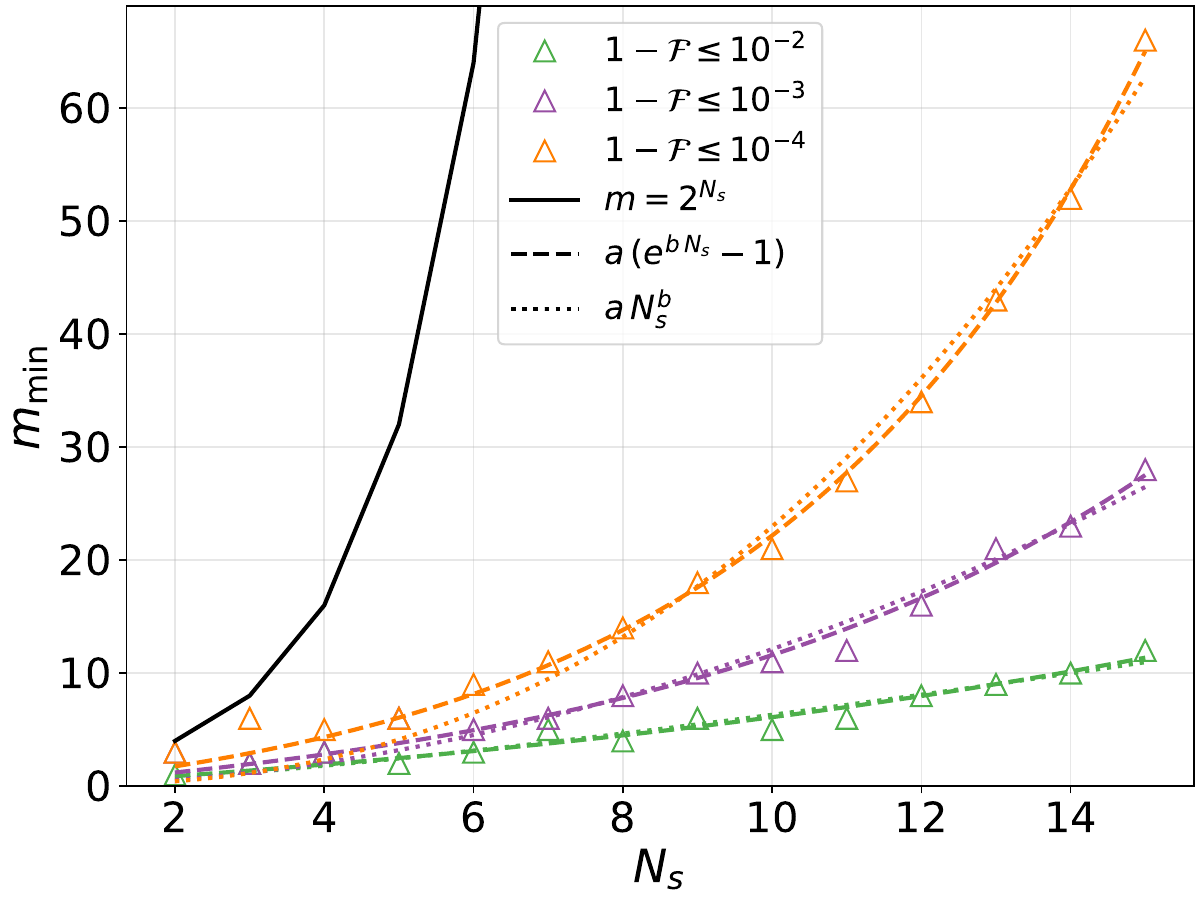}
    \caption{The scaling of $m_{\text{min}}$ required to achieve a given infidelity with $N_s$ in the paramagnetic phase. The colors indicate different infidelity thresholds, as shown in the legend.}
    \label{fig:mvsN_Para}
\end{figure}

\begin{figure}
    \centering
    \includegraphics[width=\linewidth]{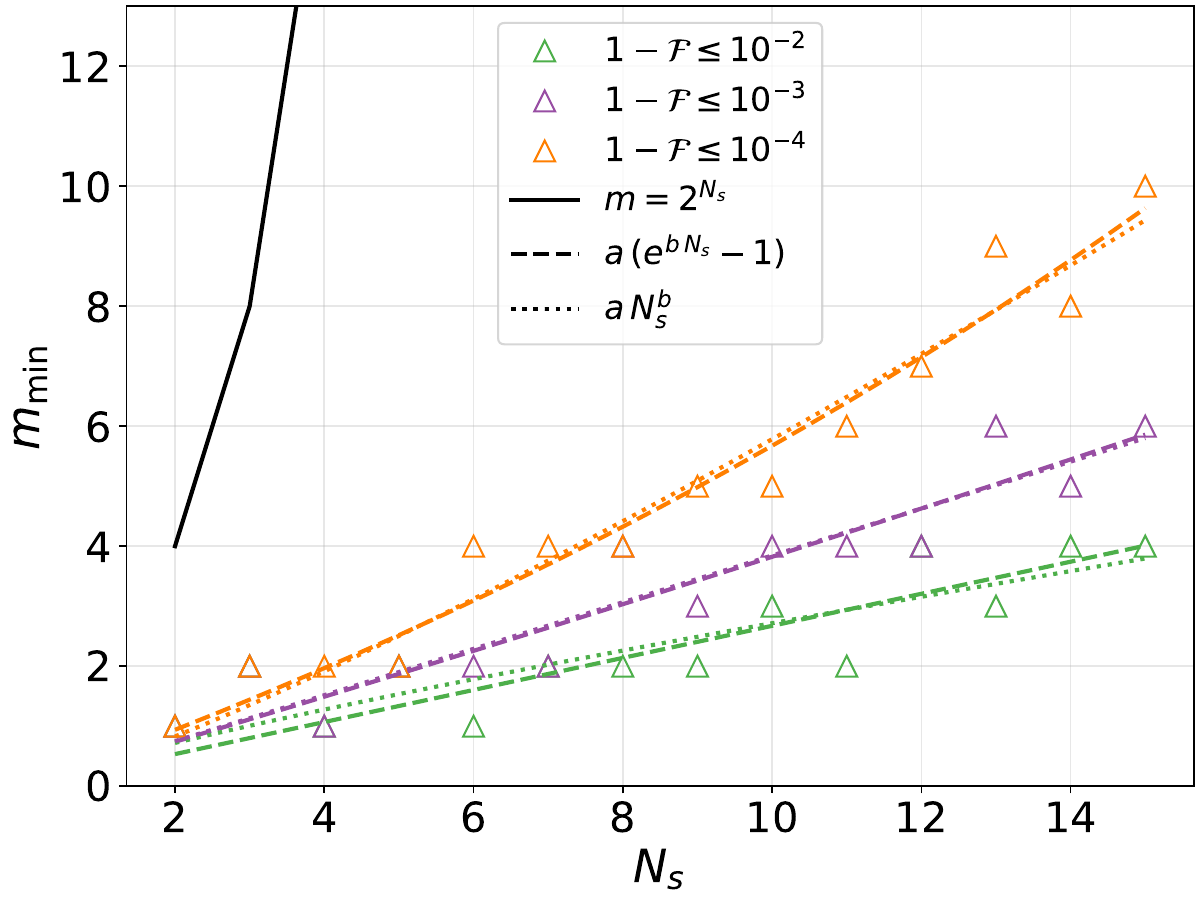}
    \caption{The scaling of $m_{\text{min}}$ required to achieve a given infidelity with $N_s$ in the antiferromagnetic phase. The colors indicate different infidelity thresholds, as shown in the legend.}
    \label{fig:mvsN_AFerro}
\end{figure}

The details and quality of the fits in Figs.~\ref{fig:mvsN_Ferro},~\ref{fig:mvsN_Para},~\ref{fig:mvsN_AFerro} of power-law and exponential functions are shown in Tab.~\ref{tab:mvsNfits}. The three sets of rows correspond to various fidelity thresholds, as shown. 

For the antiferromagnetic phase, the fits indicate a power-law/ polynomial scaling of $m_{\text{min}}$ with $N_s$. This is supported both by the quality of the power-law fits and the smallness of the exponential scaling fit parameters. For the ferromagnetic and paramagnetic phases, it is clear that for the tightest fidelity threshold (orange points), $m_{\text{min}}$ scales exponentially with $N_s$, given the difference in quality of the fits. However, for lower fidelity thresholds, the results are ambiguous given the comparable quality of the fits and, in some cases, the smallness of the exponential fit parameter. 

These scalings are determined by the spectrum of the Hamiltonian and the temperature. Specifically, the scales that are relevant are the nearest energy-level spacings and the temperature of the target Gibbs state. The diversity of scalings we observe in this work is a consequence of the diversity of the Hamiltonian's spectrum in the various phases of the XXZ model. 

\begin{table*}[ht!]
\begin{tabular}{|c|c|cc|cc|cc|}
\hline
 & & \multicolumn{2}{c|}{\textbf{Ferromagnetic}}    & \multicolumn{2}{c|}{\textbf{Paramagnetic}}   & \multicolumn{2}{c|}{\textbf{Antiferromagnetic}}    \\ \hline
 & & \multicolumn{1}{c|}{\textbf{Exp}} & \multicolumn{1}{c|}{\textbf{Pow}} & \multicolumn{1}{c|}{\textbf{Exp}} & \multicolumn{1}{c|}{\textbf{Pow}} & \multicolumn{1}{c|}{\textbf{Exp}} & \multicolumn{1}{c|}{\textbf{Pow}} \\ \hline
\parbox[t]{5mm}{\multirow{3}{*}{\rotatebox[origin=c]{90}{$\mathcal{F}\geq 99\%$}}} & $a$ & \multicolumn{1}{c|}{4.326} & \multicolumn{1}{c|}{0.2501} & \multicolumn{1}{c|}{5.426} & \multicolumn{1}{c|}{0.2719} & \multicolumn{1}{c|}{$1.803\times 10^{3}$} & \multicolumn{1}{c|}{0.4070} \\ \cline{2-8}
& $b$ & \multicolumn{1}{c|}{0.1046} & \multicolumn{1}{c|}{1.526} & \multicolumn{1}{c|}{$7.532\times 10^{-2}$} & \multicolumn{1}{c|}{1.365} & \multicolumn{1}{c|}{$1.480\times 10^{-4}$} & \multicolumn{1}{c|}{0.8244} \\ \cline{2-8}
& MSE & \multicolumn{1}{c|}{1.392} & \multicolumn{1}{c|}{2.050} & \multicolumn{1}{c|}{0.4834} & \multicolumn{1}{c|}{0.6073} & \multicolumn{1}{c|}{0.3267} & \multicolumn{1}{c|}{0.3075} \\ \hline\hline
\parbox[t]{5mm}{\multirow{3}{*}{\rotatebox[origin=c]{90}{$\mathcal{F}\geq 99.9\%$}}} & $a$ & \multicolumn{1}{c|}{42.83} & \multicolumn{1}{c|}{1.246} & \multicolumn{1}{c|}{3.636} & \multicolumn{1}{c|}{0.1431} & \multicolumn{1}{c|}{43.00} & \multicolumn{1}{c|}{0.3708} \\ \cline{2-8}
& $b$ & \multicolumn{1}{c|}{$3.803\times 10^{-2}$} & \multicolumn{1}{c|}{1.206} & \multicolumn{1}{c|}{0.1432} & \multicolumn{1}{c|}{1.928} & \multicolumn{1}{c|}{$8.516\times 10^{-3}$} & \multicolumn{1}{c|}{1.016} \\ \cline{2-8}
& MSE & \multicolumn{1}{c|}{0.3468} & \multicolumn{1}{c|}{0.1599} & \multicolumn{1}{c|}{1.0560} & \multicolumn{1}{c|}{2.014} & \multicolumn{1}{c|}{0.3094} & \multicolumn{1}{c|}{0.3112} \\ \hline\hline
\parbox[t]{5mm}{\multirow{3}{*}{\rotatebox[origin=c]{90}{$\mathcal{F}\geq 99.99\%$}}} & $a$ & \multicolumn{1}{c|}{6.427} & \multicolumn{1}{c|}{0.1482} & \multicolumn{1}{c|}{3.649} & \multicolumn{1}{c|}{$7.618\times 10^{-2}$} & \multicolumn{1}{c|}{9.886} & \multicolumn{1}{c|}{0.3588} \\ \cline{2-8}
& $b$ & \multicolumn{1}{c|}{0.1861} & \multicolumn{1}{c|}{2.387} & \multicolumn{1}{c|}{0.1956} & \multicolumn{1}{c|}{2.479} & \multicolumn{1}{c|}{$4.536\times 10^{-2}$} & \multicolumn{1}{c|}{1.207} \\ \cline{2-8}
& MSE & \multicolumn{1}{c|}{0.9163} & \multicolumn{1}{c|}{4.323} & \multicolumn{1}{c|}{1.174} & \multicolumn{1}{c|}{5.347} & \multicolumn{1}{c|}{0.2918} & \multicolumn{1}{c|}{0.3235} \\ \hline
\end{tabular}
\label{tab:mvsNfits}
\caption{Details of the power law and exponential fits to the scaling of $m$ with $N_s$.}
\end{table*}

%%%%%%%%%%%%%%%%%%%%%%%%%%%%%%%%%%%%%%%%%%%%%%%%%%%%%%%%%%%%%%%%%%%%%%%%%%%%%%%%%%%%%%%%%%

\subsection{Scaling of $\beta$ with $N_s$}

\begin{figure}[thb!]
    \centering
    \includegraphics[width=\linewidth]{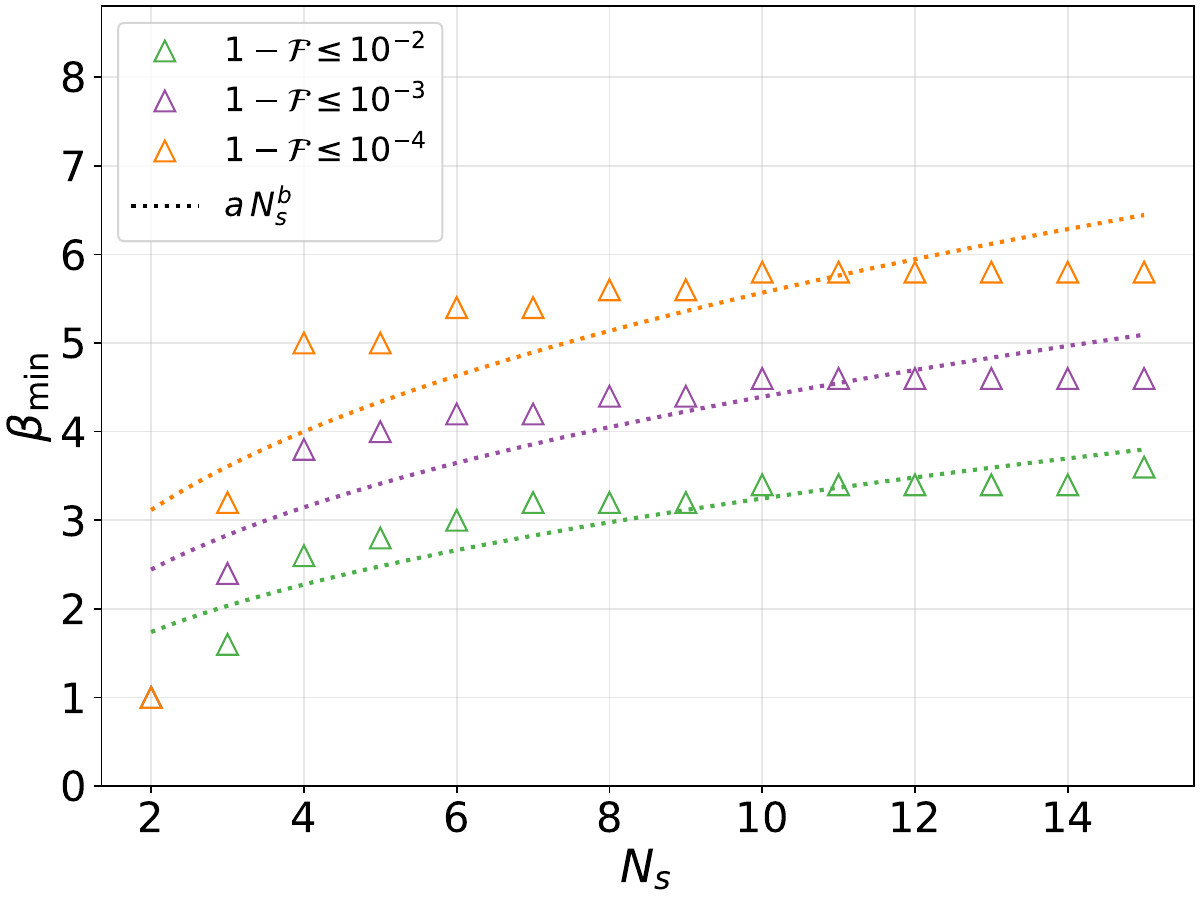}
    \caption{The scaling of the least $\beta$ required to achieve a given infidelity with $N_s$ in the ferromagnetic phase. The colors indicate different infidelity thresholds, as shown in the legend.}
    \label{fig:betavsN_Ferro}
\end{figure}

\begin{figure}[thb!]
    \centering
    \includegraphics[width=\linewidth]{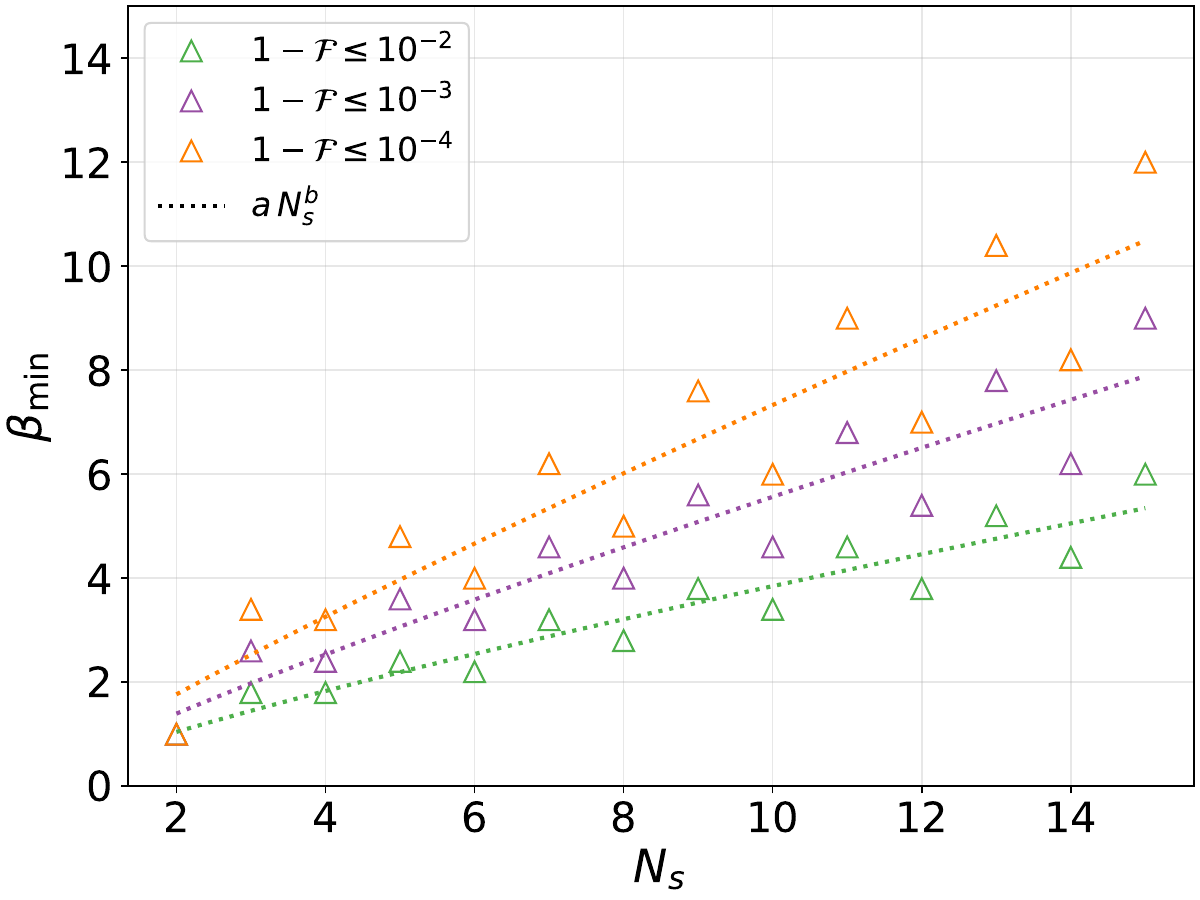}
    \caption{The scaling of the least $\beta$ required to achieve a given infidelity with $N_s$ in the paramagnetic phase. The colors indicate different infidelity thresholds, as shown in the legend.}
    \label{fig:betavsN_Para}
\end{figure}

\begin{figure}[thb!]
    \centering
    \includegraphics[width=\linewidth]{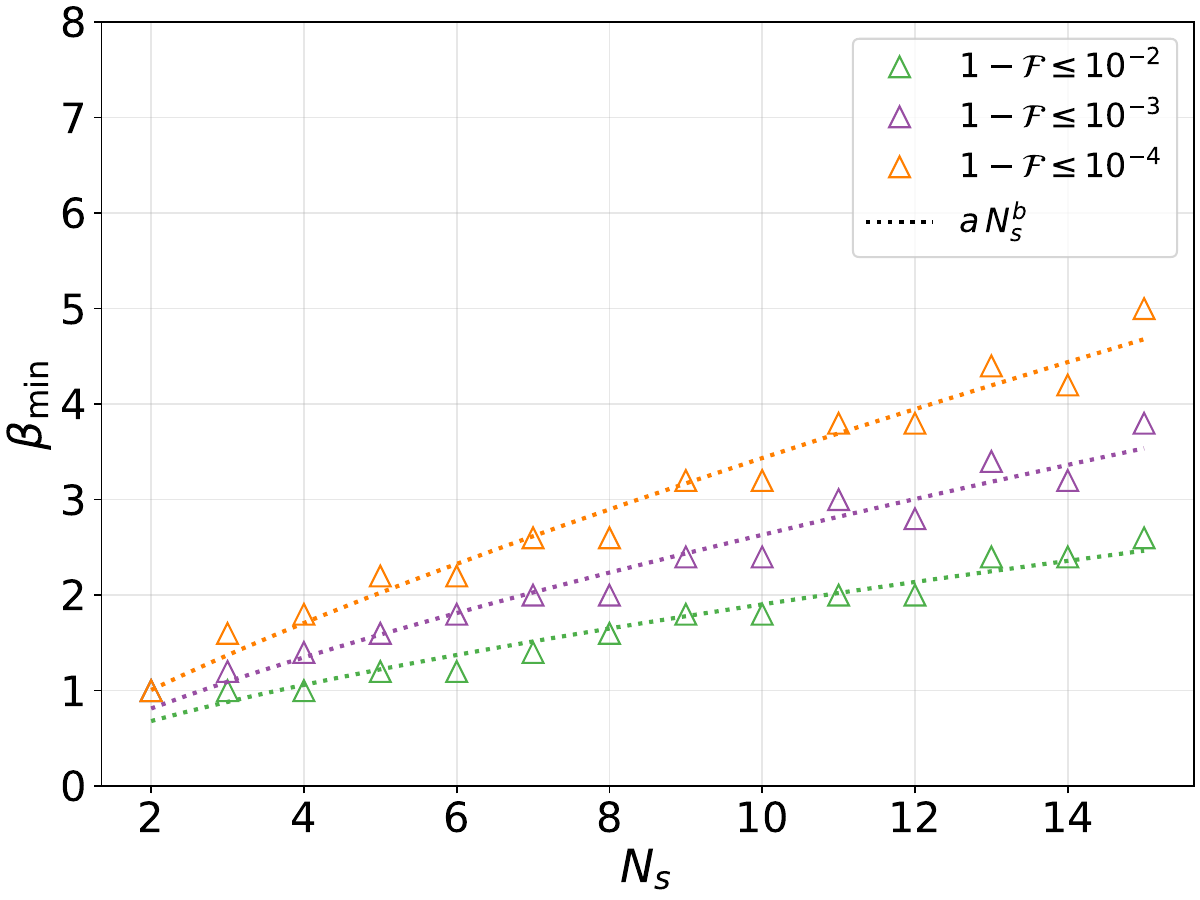}
    \caption{The scaling of the least $\beta$ required to achieve a given infidelity with $N_s$ in the antiferromagnetic phase. The colors indicate different infidelity thresholds, as shown in the legend.}
    \label{fig:betavsN_AFerro}
\end{figure}

The details and quality of the fits in Figs.~\ref{fig:betavsN_Ferro},~\ref{fig:betavsN_Para},~\ref{fig:betavsN_AFerro} of power-law functions are shown in Tab.~\ref{tab:betavsNfits}. The three sets of rows correspond to various fidelity thresholds, as shown. 

For the ferromagnetic phase, there appears to be a saturation of $\beta_{\text{min}}$ for larger $N_s$. This is because of the large gap between the ground and first excited states. As we increase $N_s$, this gap increases. As a result, a moderate value of $\beta$ is sufficient to suppress higher excited states and achieve good fidelities.

For the paramagnetic and antiferromagnetic phases, $\beta_{\text{min}}$ scales sub-linearly with $N_s$, as shown by the fit results. In both of these phases, the spectra for systems with an odd or even number of spins qualitatively differ. For instance, the model is gapless for when the number of spins is even, and has a gap that decreases with $N_s$ for an odd number of spins. This explains the alternating trend seen in Figs.~\ref{fig:betavsN_Para},~\ref{fig:betavsN_AFerro}. 

\begin{table*}[ht!]
\begin{tabular}{|c|c|c|c|c|}
\hline
 & & {\textbf{Ferromagnetic}}    & {\textbf{Paramagnetic}}    & {\textbf{Antiferromagnetic}}    \\ \hline
 \parbox[t]{5mm}{\multirow{3}{*}{\rotatebox[origin=c]{90}{$\mathcal{F}\geq 99\%$}}}
 & $a$ & 1.3296 & 0.5945 & 0.4376 \\ \cline{2-5}
 & $b$ & 0.3876 & 0.8118 & 0.6383 \\ \cline{2-5}
 & MSE & 0.1036 & 0.1794 & 0.0170 \\ \hline\hline
 \parbox[t]{5mm}{\multirow{3}{*}{\rotatebox[origin=c]{90}{$\mathcal{F}\geq 99.9\%$}}}
 & $a$ & 1.8983 & 0.7687 & 0.4902 \\ \cline{2-5}
 & $b$ & 0.3645 & 0.8596 & 0.7298 \\ \cline{2-5}
 & MSE & 0.2933 & 0.5751 & 0.0269 \\ \hline\hline
 \parbox[t]{5mm}{\multirow{3}{*}{\rotatebox[origin=c]{90}{$\mathcal{F}\geq 99.99\%$}}}
 & $a$ & 2.4273 & 0.9562 & 0.5936 \\ \cline{2-5}
 & $b$ & 0.3605 & 0.8846 & 0.7625 \\ \cline{2-5}
 & MSE & 0.5736 & 1.2084 & 0.0348 \\ \hline
\end{tabular}
\label{tab:betavsNfits}
\caption{Details of the power law and exponential fits to the scaling of $\beta$ with $N_s$.}
\end{table*}

%%%%%%%%%%%%%%%%%%%%%%%%%%%%%%%%%%%%%%%%%%%%%%%%%%%%%%%%%%%%%%%%%%%%%%%%%%%%%%%%%%%%%%%%%%
%%%%%%%%%%%%%%%%%%%%%%%%%%%%%%%%%%%%%%%%%%%%%%%%%%%%%%%%%%%%%%%%%%%%%%%%%%%%%%%%%%%%%%%%%%

\section{Paramagnetic excited states: $m=4$}\label{App:Paramagnetic3state}
In this Appendix, we investigate the case of $m=4$ in the paramagnetic phase, where TEPID-ADAPT fails to find the correct third excited state (see Fig.~\ref{fig:Infids_mscan_Para}). Instead it finds a higher excited state, whose subspace is doubly degenerate. This could be due to the optimizer getting stuck in a local minimum of the particular parametrized ansatz. Alternatively, the problem could be because the ansatz is not expressive enough to find the correct eigenspace. Here, we present evidence to disambiguate these two possibilities.

\begin{figure}[thb!]
    \centering
    \includegraphics[width=\linewidth]{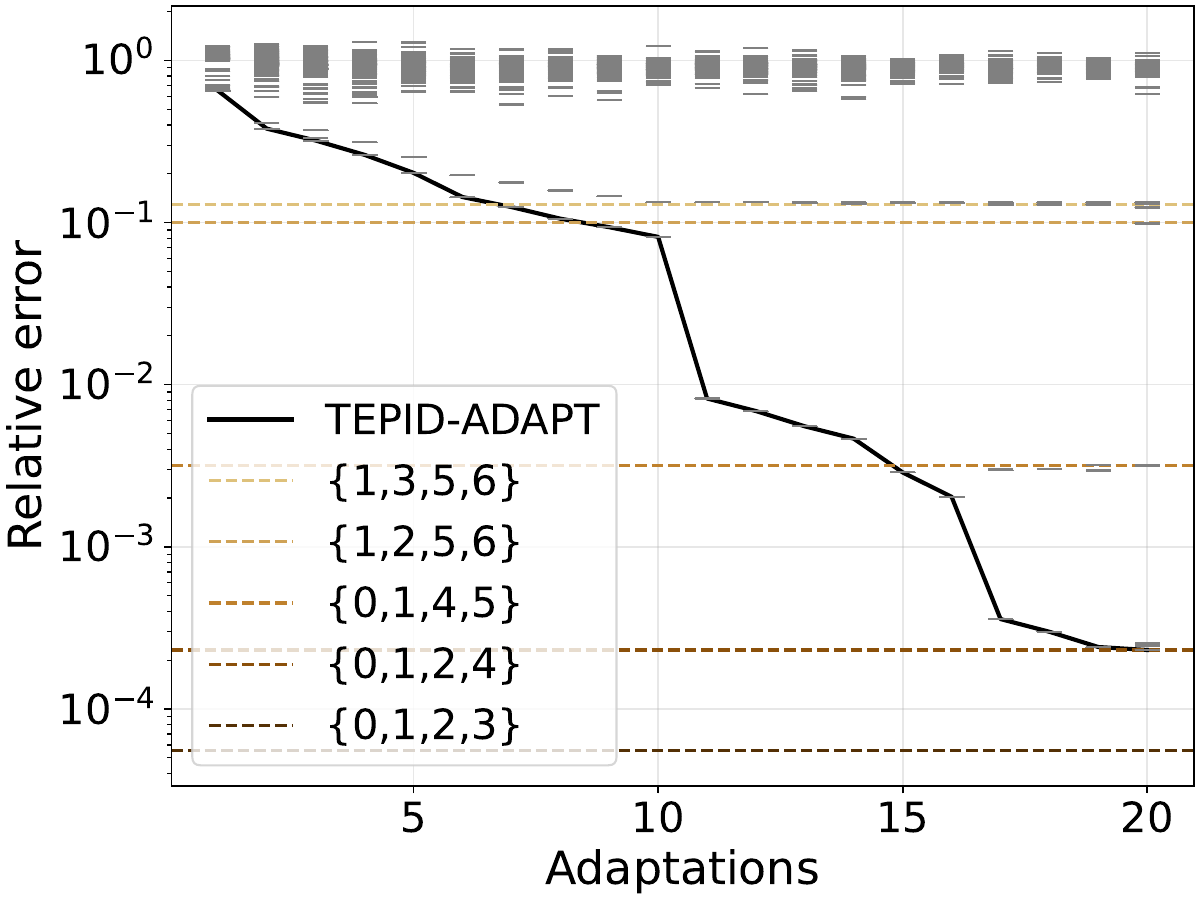}
    \caption{The relative free energy error as a function of the number of parameters in the ansatz. The solid black line is the path TEPID-ADAPT takes. The gray dashes are obtained by optimizing the corresponding ansatz after randomizing all its parameters. The different horizontal dashed lines are the relative free energy error floors of the of the indicated eigenspaces.}
    \label{fig:Randtest_Para}
\end{figure}

In ADAPT-VQE, when a new operator is added to the ansatz, the optimized parameters of the previous VQE subroutine are used as a starting point for the current optimization step. In Fig.~\ref{fig:Randtest_Para}, we show the path TEPID-ADAPT takes to convergence with the solid black line. To identify other minima of this ansatz, we randomize the parameters after the addition of each operator and optimize them, similar to Ref.~\cite{GrimsleyNPJQI2022}. We do 2500 parameter randomizations for every added operator in the ansatz. Note that not all of these optimizations have achieved convergence. Each randomization is denoted by a gray dash in Fig.~\ref{fig:Randtest_Para}. The colored dashed horizontal lines are the least achievable relative free-energy error with different eigenspaces, as indicated in the plot legend. For example, the lowest line labeled $\{0,1,2,3\}$ is the lowest relative free energy error if TEPID-ADAPT correctly finds the lowest four eigenstates of the Hamiltonian. The absence of points obtained by randomization of parameters below the solid black line is evidence that we are \textit{not} stuck in a local minimum of the ansatz. It is likely the case that the ansatz is not expressive enough to find the correct eigenspace. 

A possible solution to this is to choose a better computational basis subspace for TEPID-ADAPT initially. A different choice of an operator pool, and having tighter convergence criteria are also possible solutions.

\clearpage
\nocite{*}

\end{document}